\documentclass[twocolumn]{AASTeX7/aastex701}

\shorttitle{\texttt{harv}: fast companion orbit inference in \jax}
\shortauthors{Price-Whelan}

\usepackage{preamble/preamble-adrn}
\usepackage{listings}
\DeclareSIUnit\Mjup{M_J}

\newcommand{\harv}{\package{harv}}
\newcommand{\joker}{\textsl{The Joker}}
\newcommand{\targESS}{\ensuremath{J_{\rm eff}^*}}
\newcommand{\harvversion}{0.1.0}

\begin{document}

\title{
  \texttt{harv}:
  Fast orbit sampling for binary stars, exoplanets, and compact object companions \\
  with radial velocities and \textit{Gaia} epoch astrometry
}

\author[orcid=0000-0003-0872-7098]{Adrian M. Price-Whelan}
\email{aprice-whelan@flatironinstitute.org}
\affiliation{
  Center for Computational Astrophysics, Flatiron Institute,
  162 Fifth Ave, New York, NY 10010, USA
}

\begin{abstract}
Population studies of stellar companions (binary stars, exoplanets, compact objects)
often only use systems with well-constrained orbits.
Inferring population statistics then requires modeling the detection efficiency of the
orbit-fitting pipeline along with the survey selection function, usually through
injection/recovery tests.
An alternative approach is to include every source from the parent sample in the
population inference.
This requires generating posterior samples even for sources with sparse, noisy, or null
signals in the same manner as the well-constrained systems.
\harv\ is an open-source Python package that enables posterior sampling over orbital
parameters for radial velocity (RV) data, \gaia\ epoch astrometry data, or both, across
a range of data quality regimes.
Like its predecessor \joker, \harv\ can sample Keplerian orbit parameters either through
traditional Markov chain Monte Carlo (MCMC) or via a custom rejection sampler that
marginalizes over linear parameters in the model for efficiency.
For sparse or uninformative data, the posterior is often highly multimodal, which makes
most sampling methods inefficient and unreliable, but the rejection sampler still
returns independent posterior samples, which can then be used for hierarchical
inference.
\harv\ is built on \jax, which provides just-in-time compilation, CPU and GPU
support, and automatic differentiation.
We demonstrate \harv\ using sparse \apogee\ RVs, the planet Gaia-4b (modeling both RV
and \gaia\ \dr{4} pre-release astrometry), and a population inference with simulated RV
data.
We designed \harv\ to be flexible and fast, and to scale to the millions of sources with
time-domain data from \gaia\ \dr{4}, \sdssv, and future stellar surveys.
\end{abstract}


\section{Introduction}
\label{sec:intro}

Stars, most often, have at least one companion --- for example, another star, a brown
dwarf, planet, or compact object \citep[e.g.,][]{Duquennoy:1991, Cassan:2012, Moe:2017,
Offner:2023, Gaia:Arenou:2023, ElBadry:2024-ns}.
The distribution of companion properties (e.g., mass, orbital period, eccentricity,
etc.), their occurrence rates, and their dependence on the properties of the host star
all provide important constraints on models of star formation, binary interactions,
planet formation, and stellar evolution \citep[e.g.,][]{Fischer:2005, Sana:2012,
Duchene:2013, Moe:2019, Zhu:2021, ElBadry:2024-GaiaBinary}.
These systems are also important benchmarks for stellar astrophysics
\citep{Andersen:1991, Torres:2010, Mann:2019}.
Measuring distributions of companion properties requires time-domain survey data and
inferences of the orbital parameters for large samples of systems.

Population studies of stellar companions often start by selecting systems with
well-constrained orbits from larger catalogs.
For example, a typical pipeline will fit an orbit to every source in some survey sample,
keep the orbital solutions that pass some cuts on significance or goodness of fit, and
then use the resulting systems to infer population statistics.
The \gaia\ \dr{3} astrometric binary catalog of non-single stars (NSS) follows a similar
approach in that the pipeline decides, for each source, whether to fit a single-star or
binary-star model, and the final catalog includes only stars that pass a significance
cut on the binary model \citep{Halbwachs:2023, GaiaDR3Multiplicity}.
However, significance cuts often depend on the orbital parameters themselves, so to turn
a sample defined in this way into a robust population inference, one must model both
the parent catalog selection function \textsl{and} the detection efficiency of the
orbit-fitting-and-selection pipeline, which then requires expensive injection and
recovery tests \citep[e.g.,][]{Cumming:2008, Christiansen:2020, Rosenthal:2021,
ElBadry:2024-gaiamock}.

An alternate approach to stellar companion population inference is to instead include
all sources from the parent catalog in the model, whether or not their orbits are
precisely constrained.
In the context of a hierarchical Bayesian model (hereafter ``hierarchical model''), the
likelihood function $\mathcal{L}$ for the population parameters $\vec{\alpha}$ (e.g.,
parameters that describe the companion mass function, period distribution, etc.) follows
a hierarchical structure in which the full likelihood is a product over the individual
likelihoods for each source $n$ with data $\vec{y}_n$ and orbital parameters
$\vec{\theta}_n$ in the parent catalog:
\begin{align}
  \mathcal{L}(\vec{\alpha} \,;\, \{\vec{y}_n\}, \{\vec{\theta}_n\}) &=
    p(\{\vec{y}_n\}, \{\vec{\theta}_n\} \given \vec{\alpha}) \\
    &= \prod_n p(\vec{y}_n \given \vec{\theta}_n) \, p(\vec{\theta}_n \given \vec{\alpha})
    \quad .
\end{align}
Here the model explicitly includes the per-system orbital parameters $\vec{\theta}_n$,
so the dimensionality of the parameter space grows with the number of sources in the
parent catalog (bad).
A common approach to make the hierarchical model more tractable is to marginalize over
the per-system parameters $\vec{\theta}_n$ to instead define a marginalized likelihood
function
\begin{align}
  \mathcal{L}(\vec{\alpha} \,;\, \{\vec{y}_n\}) &=
    p(\{\vec{y}_n\} \given \vec{\alpha}) \\
  &= \prod_n
    \int \dd\vec{\theta}_n \,
      p(\vec{y}_n \given \vec{\theta}_n) \, p(\vec{\theta}_n \given \vec{\alpha})
    \quad ,
  \label{eq:pop-marginal}
\end{align}
but this integral is often not analytically solvable.
This integral can instead be approximated using a Monte Carlo sum by generating
posterior samples of the per-system parameters $\vec{\theta}_n$ for each source
independently under an ``interim'' prior $p(\vec{\theta}_n \given \vec{\alpha}_0)$, and
then reweighting the samples using the population prior $p(\vec{\theta}_n \given
\vec{\alpha})$ \citep[e.g.,][]{Hogg:2010, Foreman-Mackey:2014, Price-Whelan:2020}.
Each term in the product above is then proportional to an average of the ratio of priors
$p(\vec{\theta}_n \given \vec{\alpha}) / p(\vec{\theta}_n \given \vec{\alpha}_0)$ over
the posterior samples for source $n$ (Equation~\ref{eq:reweight} in
Section~\ref{sec:methods-rejection}).
A source with uninformative data would have posterior samples that follow the interim
prior and therefore do not constrain $\vec{\alpha}$ (or do so weakly), but importantly
the source would still appear in the product.

In this framework, we do not need to make a decision about whether a source passes some
detection threshold, and we therefore do not need to model the detection efficiency of
our orbit fitting method, which simplifies (and reduces the cost of) that aspect of the
modeling.
The challenge (and cost) here instead is that we must generate posterior samples of the
orbital parameters $\vec{\theta}_n$ for every source in the parent catalog, including
sources with sparse or noisy data, or effectively null signals.
This is a challenge for Keplerian orbit fitting: with only a few radial velocity
measurements, or astrometry at low signal-to-noise, the per-source likelihood as a
function of period can have many narrow modes separated by large regions of low
probability \citep[e.g.,][]{Price-Whelan:2017}.
Markov chain Monte Carlo (MCMC) and most sampling methods struggle in this regime.

We previously developed \joker\ \citep{Price-Whelan:2017} for this problem.
At the time, we had RV data from the Apache Point Observatory Galactic Evolution
Experiment (\apogee; \citealt{Majewski:2017}) in mind because \apogee\ often has only a
few (but precisely measured) RV measurements for hundreds of thousands of stars.
\joker\ is a custom rejection sampler that can generate independent posterior samples of
Keplerian orbital parameters for a source even when the posterior is highly multimodal.
Rejection sampling is a brute-force method that draws samples from a proposal
distribution and then accepts or rejects the samples based on the ratio of the target
pdf to the proposal pdf.
In our case, the proposal distribution is the prior pdf for the orbital parameters, and
the target pdf is the posterior pdf, so the acceptance criteria is set by the
likelihood.
Rejection sampling for posterior characterization is often considered impractical
because it scales poorly with the dimensionality of the parameter space.
With \joker, we make the problem tractable by analytically marginalizing over the linear
parameters in the Keplerian model (the velocity semi-amplitude, systemic velocity, and
sometimes extra polynomial trend parameters) to reduce the dimensionality of the
parameter space that we need to sample over (i.e., we reject samples using a marginal
likelihood).
An advantage to rejection sampling here is that the returned samples are independent
draws from the posterior, however many modes there are, so we do not need to worry about
convergence or autocorrelation analysis of chains (as we do with MCMC;
\citealt{Hogg:2018}).
\joker\ has been used in a number of population studies of stellar companions
\citep[e.g.,][]{Giesers:2019, Price-Whelan:2020, Rosenthal:2021}.

We redesigned and re-implemented the core ideas of \joker\ from the ground up with two
contemporary stellar surveys in mind, \gaia\ \dr{4} and the Milky Way Mapper (MWM)
program of \sdssv\ \citep{Kollmeier:2026}.
\gaia\ \dr{4}, which will be released on 2026-12-02, will include a wealth of
time-series data, including usable (for companion inference) epoch astrometry and radial
velocities for tens to hundreds of millions of stars.
MWM is continuing a stellar survey with the \apogee\ spectrograph and will provide
precise multi-epoch RVs for millions of stars, most of which will have only a few
visits.
For many of these sources, we will have both RVs and epoch astrometry, and to make the
most of the full samples we will need to be able to model the companions of faint and
therefore low signal-to-noise host stars.
With these surveys, we will need to generate posterior samples efficiently for
millions of sources, which is prohibitive with the current \joker\ implementation.

Here we present \harv: a successor to \joker\ and full re-implementation built on \jax\
\citep{Bradbury:2018} and \package{numpyro} \citep{phan2019composable, bingham2019pyro}.
\harv\ is faster (see Section~\ref{sec:performance}) and more flexible than the original
\joker\ implementation, and works with both RVs and \gaia\ epoch astrometry (with
support for more data types to come).
Because it is built on \jax, we can compile and use the same model code on CPUs or GPUs,
and it is differentiable, so it can be used in gradient-based MCMC samplers or
variational inference methods instead of the built-in rejection sampler.

Many packages and methods exist to sample or fit Keplerian orbits to data, including
packages for RV data (\texttt{juliet} \citealt{Espinoza:2019};
  \texttt{RadVel} \citealt{Fulton:2018}),
for astrometry and combinations of data (\texttt{exoplanet}
\citealt{Foreman-Mackey:2021};
  \texttt{ExoPlanetLab} \citealt{Hobbs:2026};
  \texttt{Exo-Striker} \citealt{Trifonov:2019};
  \texttt{kima} \citealt{Faria:2018};
  \texttt{Octofitter} \citealt{Thompson:2023};
  \texttt{orbitize!} \citealt{Blunt:2020};
  \texttt{orvara} \citealt{Brandt:2021}),
and for forward modeling \gaia\ epoch astrometry and the \dr{3} orbit-fitting pipeline
(\texttt{gaiamock}, \citealt{ElBadry:2024-gaiamock}).
Most of these packages are designed mainly to characterize individual systems in which a
companion has been confidently detected, using MCMC or nested sampling.
Some of these packages also include samplers designed for poorly constrained or
multimodal posterior pdfs, such as OFTI in \texttt{orbitize!} \citep{Blunt:2017},
diffusive nested sampling (DNest4; \citealt{Brewer:2014}) in \texttt{kima}
\citep{Faria:2018}, and non-reversible parallel tempering (\texttt{Pigeons.jl}) in
\texttt{Octofitter} \citep{Surjanovic:2023}.
Recent work has used \texttt{Octofitter} and \texttt{Pigeons.jl} to compute companion
posterior pdfs for about $10^5$ stars in the \textit{Hipparcos}--\gaia\ G23H catalog
\citep{Thompson:2026a} and to infer the companion population from a subset of these
stars \citep{Thompson:2026b}.
Like this work, \harv\ is designed for population inference with large samples of
sources with companions.

\section{Software design}
\label{sec:design}

\begin{table*}
  \centering
  \caption{
    The parameterizations implemented in \harv, with their nonlinear and linear
    parameters.
    For \texttt{SqrtEccRV}, $\lambda = \omega - 2\pi\tau$ is the mean longitude at the
    reference time $t_{\rm ref}$.
    For the Fourier parameterizations, the number of Fourier terms is user-defined, so
    the number of linear coefficients $a_k$ and $b_k$ depends on the number of terms.
    \label{tab:parameterizations}
  }
  \begin{tabular}{lll}
    \hline
    Parameterization & Nonlinear parameters & Linear parameters \\
    \hline
    \multicolumn{3}{l}{\textit{Radial velocities}} \\
    \texttt{StandardRV} &
      $P$, $e$, $\tau$, $\omega$ &
      $K$, $v_0$ \\
    \texttt{SqrtEccRV} &
      $P$, $\sqrt{e}\cos\omega$, $\sqrt{e}\sin\omega$, $\lambda$ &
      $K$, $v_0$ \\
    \texttt{FourierRV} &
      $P$ &
      $v_0$, $a_k$, $b_k$ \\
    \multicolumn{3}{l}{\textit{\gaia\ epoch astrometry}} \\
    \texttt{StandardGaiaAstrometry} &
      $P$, $e$, $\tau$, $\omega$, $\Omega$, $\cos i$ &
      $\alpha_0$, $\delta_0$, $\mu_\alpha$, $\mu_\delta$, $\varpi$, $a_0$ \\
    \texttt{ThieleInnesGaiaAstrometry} &
      $P$, $e$, $\tau$ &
      $\alpha_0$, $\delta_0$, $\mu_\alpha$, $\mu_\delta$, $\varpi$, $A$, $B$, $F$,
      $G$ \\
    \texttt{FourierGaiaAstrometry} &
      $P$ &
      $\alpha_0$, $\delta_0$, $\mu_\alpha$, $\mu_\delta$, $\varpi$, $a_k$, $b_k$ \\
    \hline
  \end{tabular}
\end{table*}

\harv\ is designed to generate posterior samplings over the orbital parameters for
two-body systems.
Much of the design of \harv\ is informed by the fact that, for such systems, the model
prediction for the observed data is a nonlinear function of some of the parameters and a
linear function of the others.
That is, for a source with $N$ measurements $\vec{y}$, we can write the prediction as
\begin{equation}
  \vec{y} = \mat{X}(\vec{\theta})\,\vec{\beta} + \mathrm{noise}
  \label{eq:linear-model}
\end{equation}
where the $N \times Q$ design matrix $\mat{X}$ is, in general, a function of the
nonlinear parameters $\vec{\theta}$ and $\vec{\beta}$ is a length-$Q$ vector of the
linear parameters.
The implication of this structure is that, if we adopt conjugate priors for the linear
parameters, we can analytically marginalize over the linear parameters to reduce the
dimensionality of the parameter space we need to sample over.
We describe this marginalization in more detail in Section~\ref{sec:methods-marginal}.
Table~\ref{tab:notation} (Appendix~\ref{apx:conventions}) lists the symbols that we use
in more than one section of the main text.

\subsection{Parameterizations of the models}
\label{sec:design-parameterizations}

\harv\ supports several parameterizations of the models that predict the data.
For example, for radial velocity data, the standard parameterization of the Keplerian
model in \harv\ is
\begin{equation}
  v(t) = v_0 + K\,[\cos(\omega + f(t \,;\, P, e, \tau)) + e\,\cos\omega]
  \quad ,
  \label{eq:rv-model}
\end{equation}
where $v_0$ is the systemic velocity, $K$ is the velocity semi-amplitude, $e$ is the
eccentricity, $\omega$ is the argument of pericenter, and the true anomaly $f$ depends
on time $t$ with the period $P$, the eccentricity, and the orbital phase $\tau$.
We specify the phase $\tau$ as the time of pericenter passage in units of the period,
$\tau = (t_{\rm peri} - t_{\rm ref}) / P$, relative to a reference time $t_{\rm ref}$.
Of these parameters, four ($P$, $e$, $\omega$, $\tau$) are nonlinearly related to the
prediction $v(t)$, and two ($K$, $v_0$) are linear.
This is implemented in \harv\ as the \texttt{StandardRV} parameterization class.
But we also support alternate parameterizations of the Keplerian model.
For example, \harv\ supports a Fourier series representation of the Keplerian model,
which is linear in all parameters (coefficients) except the period, and is implemented
as the \texttt{FourierRV} parameterization class, among others (see
Table~\ref{tab:parameterizations}).
In the code, a parameterization specifies the names and units of the parameters of a
model, along with determining which parameters are linear, how the design matrix is
computed from the nonlinear parameters, and a default prior probability density function
(pdf) for all parameters.

As another example, for \gaia\ epoch astrometry, the data are one-dimensional
``along-scan'' positions of the source \citep{Lindegren:2012}, which are a projection of
the two-dimensional sky position onto the scan direction of the satellite over time.
We model the along-scan position $y_{\rm AL}$ at time $t$ as the sum of the standard
five-parameter astrometric model and a Keplerian orbit of the photocenter
\citep[e.g.,][]{Halbwachs:2023}.
In terms of the Thiele--Innes constants $(A, B, F, G)$ and the astrometric in-plane
coordinates $(X, Y)$ of the orbit (see Appendix~\ref{apx:conventions}), the along-scan
position is
\begin{align}
  y_{\rm AL}(t) =
    &\left[\alpha_0 + \mu_\alpha\,\Delta t + B\,X + G\,Y\right] \, \sin\psi \nonumber \\
    + &\left[\delta_0 + \mu_\delta\,\Delta t + A\,X + F\,Y\right] \, \cos\psi \nonumber \\
    + &\varpi\,H_\varpi
  \quad ,
  \label{eq:al-model}
\end{align}
where $(\alpha_0, \delta_0)$ are offsets in position at the reference time relative to a
reference position, $(\mu_\alpha, \mu_\delta)$ are the proper motion components,
$\varpi$ is the parallax, $\Delta t = t - t_{\rm ref}$, and $\psi$ and $H_\varpi$ are
the scan angle and the along-scan parallax factor of the transit (both of which are
provided as metadata in \gaia\ \dr{4}).
The orbit enters through the astrometric coordinates $(X, Y)$
(Equation~\ref{eq:conv-orbit}) and through the Thiele--Innes constants $(A, B, F, G)$
(Equation~\ref{eq:conv-thiele-innes}), which include the angular semi-major axis of the
photocenter orbit, $a_0$, multiplied by functions of $\omega$, the longitude of the
ascending node $\Omega$, and the inclination $i$ (i.e. the orientation angles of the
orbit).
The prediction is linear in the five astrometric parameters ($\alpha_0$, $\delta_0$,
$\mu_\alpha$, $\mu_\delta$, $\varpi$).
At fixed orientation of the orbit $(\omega, \Omega, i)$, it is also linear in the
angular semi-major axis $a_0$, and it is linear in the Thiele--Innes constants
themselves.
We therefore support both options: a parameterization class that uses the Thiele--Innes
constants as linear parameters, \texttt{ThieleInnesGaiaAstrometry}, and a
parameterization class that uses the orientation angles and angular semi-major axis as
nonlinear parameters, \texttt{StandardGaiaAstrometry}.
We list our conventions for the orientation of the orbit in
Appendix~\ref{apx:conventions}, which are chosen so that the same ($\omega$, $\Omega$,
$i$) describe RV data and astrometry consistently for a star.

Table~\ref{tab:parameterizations} lists the parameterizations that are currently
supported.
The user then chooses the prior pdfs, which determine whether the linear parameters can
be marginalized over, and chooses the inference method (rejection sampling, MCMC, etc.).
A user can therefore add a new data format by writing a parameterization class that
computes the model prediction, and then use the rest of \harv\ to set prior pdfs,
marginalize (or not) over the linear parameters, and generate posterior samples.

\subsection{Assumptions}
\label{sec:design-assumptions}

As with any statistical model, we make a number of assumptions in \harv\ about the data
and the underlying physical systems we are modeling.
For this version of \harv, we assume:
\begin{enumerate}
  \item that each source has one unseen companion and that the motion is well described
  by the two-body problem,\footnote{Our assumption is actually slightly weaker, as we do
  support polynomial trends in the model, which can absorb the acceleration from a
  long-period outer companion to an inner companion that is the target of the main
  model. See Section~\ref{sec:design-extensions}.}
  \item that the measurement times are known and specified in a barycentric reference
  frame,
  \item and that the noise is Gaussian with known variances, possibly inflated by
  additional parameters (see below).
\end{enumerate}
For astrometry, we additionally assume that the measured position is that of the
photocenter of the single luminous object.
We return to these assumptions in Section~\ref{sec:discussion}.

\subsection{Model extensions}
\label{sec:design-extensions}

Real data often violate the above assumptions.
For example, many stars have more than one companion.
Some of these cases are hierarchical systems with an inner (short-period) companion and
an outer (long-period) companion, which we do partially support.
Or, the noise model may be wrong: there may be correlated noise (instrumental or
astrophysical) in the data, or systematic offsets between instruments, or the reported
uncertainties may be underestimated.
\harv\ supports extending the base models with additional components and parameters to
account for these effects.

As they are currently implemented, extensions can be one of two kinds: (1) extensions
that add columns to the design matrix (i.e., add linear parameters), or (2) extensions
that modify the noise covariance matrix of the data.
Examples of the first kind of extension include a constant velocity offset for each
instrument when the data come from RV programs with different calibrations, or a
polynomial trend in time to absorb the acceleration from a long-period companion.
The model prediction is linear in these parameters, so \harv\ can optionally marginalize
over them, like the other linear parameters of the Keplerian part of the model.
Examples of the second kind of extension are either an extra ``jitter'' or ``excess
variance'' term that is added to the variance of each measurement, or a Gaussian process
(GP) model for correlated noise in the data.
We currently support GP models via the \package{tinygp} package
\citep{Foreman-Mackey:2026}, which is also built on \jax.
For both a jitter term and a GP model, the extra parameters of the noise model enter the
likelihood nonlinearly, so \harv\ cannot analytically marginalize over these additional
parameters.

\subsection{Implementation details}
\label{sec:design-implementation}

\harv\ models are built on \jax\ \citep{Bradbury:2018}.
In a break from many packages in the \jax\ ecosystem, we require that all arrays and
calculations use double precision (\texttt{float64}).
We use \package{numpyro} \citep{phan2019composable} to represent and reason about prior
pdfs and to run MCMC or other inference methods, \package{equinox} \citep{Kidger:2021}
to define models, priors, and data containers as immutable dataclasses that \jax\ can
trace through and transform, and \package{unxt} \citep{Starkman:2025} so that data,
parameters, priors, and samples carry physical units.
When needed, we compute the eccentric anomaly with a \jax-compatible implementation of
the Kepler solver from \package{orvara} \citep{Brandt:2021, Raposo-Pulido:2017}.
The log-probability of a model is a pure function of the parameter values and can be
extracted from the \package{numpyro} model, so this can also be passed to other
libraries in the \jax\ ecosystem; we show an example of this in
Section~\ref{sec:cases-interop} where we demonstrate using \harv\ with
\package{blackjax} \citep{Cabezas:2024} to run a nested sampling with a \harv\ model.

\section{Methods}
\label{sec:methods}

The main goal of \harv\ is to enable generating posterior samples of orbital parameters
for large numbers of sources even when the posterior pdf is highly multimodal or
structured.
As with \joker\ \citep{Price-Whelan:2017}, the main trick we use to make this more
tractable is to analytically marginalize over the linear parameters in the model, which
reduces the dimensionality of the parameter space that we need to sample over.
We then also implement a custom rejection sampler that can efficiently sample from the
remaining nonlinear parameters.
In this section, we describe the marginalization and rejection sampling methods
implemented in \harv.

\subsection{Marginalizing the linear parameters}
\label{sec:methods-marginal}

As in Section~\ref{sec:intro}, we assume that we have data $\vec{y} = y_n$ for
$n=1,\ldots,N$ measurements of the radial velocity, along-scan position, or both for a
star with a companion whose gravitational influence on the observed star follows
Keplerian motion.
We assume that the data are Gaussian distributed around the model prediction with
covariance matrix $\mat{C}$, and that we know the measurement times $\vec{t} = t_n$
exactly in a barycentric reference frame.
The data covariance matrix may be diagonal and fixed if the observations are independent
and the uncertainties are known, or it may include terms that depend on parameters of
the model, such as a jitter (white noise) term $s^2\,\identity$ with $\identity$ as the
$N \times N$ identity matrix, or a Gaussian process model for correlated noise
$\mat{\Sigma}_\mathrm{GP}$.
The model prediction is a linear function of some parameters $\vec{\beta}$ and a
nonlinear function of the remaining parameters $\vec{\theta}$, as in
Equation~\ref{eq:linear-model}, where $\mat{X}(\vec{\theta})$ is the $N \times Q$ design
matrix (which depends on $\vec{\theta}$) and $\vec{\beta}$ contains the $Q$ linear
parameters (Table~\ref{tab:parameterizations}), so that the likelihood is a Gaussian,
\begin{equation}
  p(\vec{y} \given \vec{\theta}, \vec{\beta}) =
    \norm(\vec{y} \given \mat{X}(\vec{\theta})\,\vec{\beta}, \mat{C}) \quad .
\end{equation}
If the prior pdf over the linear parameters is also Gaussian (and potentially dependent
on the nonlinear parameters $\vec{\theta}$),
\begin{equation}
  p(\vec{\beta} \given \vec{\theta}) =
    \norm(\vec{\beta} \given \vec{\mu}(\vec{\theta}), \mat{\Lambda}(\vec{\theta}))
    \quad ,
\end{equation}
then we can refactor the product of the likelihood and this prior into a product of two
different Gaussians \citep{Hogg:2020},
\begin{align}
  p(\vec{y} \given \vec{\theta}, \vec{\beta}) \, p(\vec{\beta} \given \vec{\theta}) &=
    \norm(\vec{y} \given \mat{X}\,\vec{\beta}, \mat{C})\,
      \norm(\vec{\beta} \given \vec{\mu}, \mat{\Lambda}) \\
  &= \norm(\vec{y} \given \mat{X}\,\vec{\mu},
      \mat{C} + \mat{X}\,\mat{\Lambda}\,\transp{\mat{X}})\,
    \norm(\vec{\beta} \given \vec{a}, \mat{A})
  \label{eq:refactor} \\
  \mat{A}^{-1} &= \mat{\Lambda}^{-1} + \transp{\mat{X}}\,\mat{C}^{-1}\,\mat{X} \\
  \vec{a} &= \mat{A}\,(\mat{\Lambda}^{-1}\,\vec{\mu}
    + \transp{\mat{X}}\,\mat{C}^{-1}\,\vec{y})
  \quad .
\end{align}
The first term on the right-hand side of Equation~\ref{eq:refactor} only depends on the
data and the nonlinear parameters --- this is the marginal likelihood of the data given
the nonlinear parameters,
\begin{equation}
  p(\vec{y} \given \vec{\theta})
    = \norm(\vec{y} \given \mat{X}\,\vec{\mu},
      \mat{C} + \mat{X}\,\mat{\Lambda}\,\transp{\mat{X}})
  \quad .
  \label{eq:marginal-likelihood}
\end{equation}
The second factor is the \textit{posterior} pdf of the linear parameters conditioned on
the nonlinear parameters, and this is also a Gaussian with mean $\vec{a}$ and covariance
$\mat{A}$.
For brevity, we write the marginal likelihood as $\mathcal{G}(\vec{\theta}) \equiv
p(\vec{y} \given \vec{\theta})$, and $\mathcal{G}_j = \mathcal{G}(\vec{\theta}_j)$ for a
particular sample $\vec{\theta}_j$.
This is the marginalization that \joker\ uses (see Appendix A of
\citealt{Price-Whelan:2020}), but \harv\ generalizes the implementation to support
additional columns in the design matrix (see Section~\ref{sec:design-extensions}), a
nonzero prior mean vector $\vec{\mu}$, a prior covariance matrix $\mat{\Lambda}$ that
can depend on the nonlinear parameters $\vec{\theta}$, and a dense data covariance
matrix $\mat{C}$.
In code, when $\mat{C}$ is diagonal, we do not construct the $N \times N$ covariance
matrix in Equation~(\ref{eq:marginal-likelihood}) and we instead use the matrix
inversion and determinant lemmas to improve the evaluation cost of the marginal
likelihood (which is the main computational bottleneck).

The marginalization above requires that the prior pdf over each linear parameter is
conjugate to the likelihood, which here means the prior pdf must be a normal
distribution, or, new to \harv, a truncated normal distribution, or a mixture of these
(see below).
In code, if a user specifies a prior pdf for a linear parameter that is not a normal,
truncated normal, or mixture of these, this is allowed, but then \harv\ samples that
linear parameter explicitly along with the nonlinear parameters.

We support truncated normal distributions because some parameters may have physical
constraints on their allowed values.
For example, a true parallax must be positive, or in a double-lined ``SB2'' binary, the
two stars must have opposite signs of the velocity semi-amplitude.
For these cases we support marginalization over truncated normal prior pdfs --- that is,
a normal distribution that is zero outside an allowed domain $\mathbb{S}$ (a lower
bound, an upper bound, or both, for each truncated parameter).
We can still marginalize analytically because the truncation does not change the
refactorization in Equation~(\ref{eq:refactor}); it only restricts the domain of the
integral over $\vec{\beta}$ to $\mathbb{S}$.
The first term on the right-hand side of Equation~(\ref{eq:refactor}) does not depend on
$\vec{\beta}$, so it can come out of the integral, and the marginal likelihood
(Equation~\ref{eq:marginal-likelihood}) becomes
\begin{equation}
  p(\vec{y} \given \vec{\theta}) =
    \frac{Z_{\rm post}}{Z_{\rm prior}} \,
    p_{\rm untrunc}(\vec{y} \given \vec{\theta})
  \quad ,
  \label{eq:truncated}
\end{equation}
where $p_{\rm untrunc}(\cdot)$ is Equation~(\ref{eq:marginal-likelihood}) evaluated with
the untruncated prior, and
\begin{align}
  Z_{\rm prior} &= \int_\mathbb{S} \dd\vec{\beta} \,
    \norm(\vec{\beta} \given \vec{\mu}, \mat{\Lambda})
    \label{eq:truncated-Zprior} \\
  Z_{\rm post} &= \int_\mathbb{S} \dd\vec{\beta} \,
    \norm(\vec{\beta} \given \vec{a}, \mat{A})
  \label{eq:truncated-Zpost}
\end{align}
are the integrals of the untruncated prior pdf and of the untruncated conditional
posterior pdf over the domain $\mathbb{S}$.

With independent prior pdfs over the linear parameters, $\mat{\Lambda}$ is diagonal, so
$Z_{\rm prior}$ (Equation~\ref{eq:truncated-Zprior}) is a product of one-dimensional
integrals that we evaluate with the normal cumulative distribution function (cdf).
$Z_{\rm post}$ (Equation~\ref{eq:truncated-Zpost}) is harder: the term
$\transp{\mat{X}}\,\mat{C}^{-1}\,\mat{X}$ in $\mat{A}^{-1}$ makes the linear parameters
covariant in the conditional posterior pdf, so $Z_{\rm post}$ is the integral of a
multivariate Gaussian with a dense covariance matrix over a rectangular domain.
The parameters that are not truncated can be integrated out, which leaves a Gaussian in
only the truncated parameters.
For one truncated parameter, $Z_{\rm post}$ is a difference of two values of the normal
cdf.
For two truncated parameters, it is a sum over the four corners of the domain (we give
these expressions, and how we evaluate them and sample from the truncated conditional
posterior pdf, in Appendix~\ref{apx:truncated}).
\harv\ does not support more than two truncated parameters because there is no
comparably simple expression.

We also support prior pdfs that are a mixture of normal distributions, which are useful
for approximating any other prior pdf shape (e.g., power laws, lognormals, etc.).
The expressions above generalize to also support prior pdfs that are mixtures of normal
distribtions, $p(\vec{\beta} \given \vec{\theta}) \propto \sum_i \eta_i\,
\norm(\vec{\beta} \given \vec{\mu}_i, \mat{\Lambda}_i)$ with $\sum_i \eta_i = 1$.
The integrals over $\vec{\beta}$ then become weighted sums over the mixture components,
\begin{equation}
  p(\vec{y} \given \vec{\theta}) =
    \frac{\sum_i \eta_i\,Z_{{\rm post},i}\,
      p_{{\rm untrunc},i}(\vec{y} \given \vec{\theta})}
    {\sum_i \eta_i\,Z_{{\rm prior},i}}
  \quad ,
  \label{eq:mixture}
\end{equation}
where the subscript $i$ indicates a quantity computed for mixture component $i$, and all
mixture components share the same truncated domain.

\subsection{Priors that depend on the nonlinear parameters}
\label{sec:methods-priors-nonlinear}

Using normal prior pdfs for the linear parameters is convenient because it allows us to
analytically marginalize them, but unconditional normal priors can lead to unphysical or
unrealistic prior beliefs about the systems being modeled.
As an example, consider the velocity semi-amplitude $K$ of a Keplerian orbit, which is a
linear parameter in the radial velocity model (Equation~\ref{eq:rv-model}).
A normal prior pdf on $K$ that is constant with period $P$ implies that the prior pdf on
the companion mass $m_2$ is a function of period, which is probably not a good choice in
general \citep{Price-Whelan:2020}.
Instead, we can pick a prior scale that is fixed in some other physical quantity, like
the companion mass or the physical size of the orbit.
In \citet{Price-Whelan:2020}, we defined a scale for the prior on $K$ that is fixed in
companion mass, where the variance of the prior pdf is
\begin{equation}
  \sigma_K^2 = \sigma_{K,0}^2\,\left(\frac{P}{P_0}\right)^{-2/3}\,(1 - e^2)^{-1}
  \quad ,
  \label{eq:sigma-K}
\end{equation}
$\sigma_{K,0}$ is a velocity scale, and $P_0$ is a reference period.
At fixed primary mass, this is fixed in companion mass, which is a more physically
motivated choice.

We can make a similar argument for the prior on the angular semi-major axis of the
photocenter orbit, $a_0$, in astrometric models.
If we adopt a prior on $a_0$ that is the same at all periods and distances, then we are
implicitly assuming that the physical size of the orbit is a function of period and
distance, which is probably not a good assumption.
For astrometric models we therefore adopt a zero-mean normal prior on $a_0$ and on
the proper motion components with standard deviations
\begin{equation}
  \sigma_{a} = \sigma_{a,0}\,\left(\frac{P}{P_0}\right)^{2/3}\,\varpi
  \quad \mathrm{and} \quad
  \sigma_\mu = \sigma_{v,0}\,\varpi
  \quad ,
  \label{eq:sigma-a}
\end{equation}
where $\sigma_{a,0}$ is a length scale (e.g., in \unit{au}) and
$\sigma_{v,0}$ is a velocity scale, so that the priors are fixed in physical size and in
tangential velocity.
These expressions depend on the parallax $\varpi$, which is itself a linear parameter
of the astrometric model.
When using these priors, we therefore technically have to sample the parallax explicitly
along with the nonlinear parameters.
This lowers the efficiency of the rejection sampler for every source, because the
parallax is usually measured \emph{much} more precisely than the width of the prior on
parallax.
When the parallax is already well measured, for example in a catalog like \gaia, we
can instead compute the prior widths using the catalog value of the parallax.
We can then treat the parallax as a linear parameter to marginalize over.
This is technically wrong, because (a) the parallax itself may be biased by the orbit,
and (b) the prior widths will be biased by the fractional error of the fixed parallax.
However, in practice this is a small effect because the prior widths are usually much
larger than the fractional error of the parallax --- the parallax is still handled
properly as a parameter in the model, so this only concerns the scale of the priors over
the \emph{other} linear parameters.
When the parallax is poorly measured, however (e.g., $\sigma_\varpi/\varpi \gtrsim
0.1$), we recommend sampling the parallax explicitly, because the prior widths will then
be more accurate and the additional cost of sampling the parallax is small when it
is not well constrained.
In either case, the posterior parallax should be compared with the value used for the
prior widths, and if they differ by more than a few percent, the fit can be iterated
with the posterior value.

\subsection{The rejection sampler}
\label{sec:methods-rejection}

\harv\ models are implemented on top of \package{numpyro} so that prior pdfs are
specified as \package{numpyro} distribution objects and the likelihood is a
\jax-compatible function of the parameters and data.
This means that we can use any of the inference methods implemented in \package{numpyro}
to generate posterior samples of the parameters.
However, for all of the reasons described in Section~\ref{sec:intro}, we expect that the
posterior pdfs for many sources may be highly multimodal and structured, and so we also
provide a custom rejection sampler that is designed specifically to handle these cases.

The core algorithm of the rejection sampler is the same as in \joker\
\citep{Price-Whelan:2017}:
\begin{enumerate}
  \item Generate $J$ samples $\vec{\theta}_j$ from the prior pdf over the nonlinear
    parameters (and over any linear parameters that are explicitly sampled, i.e., not
    marginalized).
  \item Evaluate the marginal likelihood $\mathcal{G}_j = \mathcal{G}(\vec{\theta}_j)$
  for each sample (see Section~\ref{sec:methods-marginal}).
  \item Find the maximum value $\mathcal{G}_{\rm max}$ across all of the samples.
  \item For each sample $j$, choose a random number $u_j$ uniformly between 0 and
    $\mathcal{G}_{\rm max}$, and reject the sample if $\mathcal{G}_j < u_j$.
  \item For each of the $S$ samples that survive, generate a sample of the linear
    parameters from the conditional posterior pdf,
    $\norm(\vec{\beta} \given \vec{a}_j, \mat{A}_j)$.
\end{enumerate}
If the sampling of the prior pdf is dense enough (or the data are uninformative), many
samples will survive the rejection step.
The surviving samples are, by construction, independent samples from the posterior pdf.
This is a brute force method, and for many sources, most of the computational effort
will be spent evaluating the marginal likelihood for samples that are ultimately
rejected.
But the benefit is that there is no burn-in, no need to compute autocorrelation times,
no need to tune proposal distributions, and no need to worry about convergence of
MCMC chains.
The likelihood evaluations are also embarrassingly parallel, so the method parallelizes
easily across multiple CPU cores or GPUs.
We do, however, have to choose the number of prior samples $J$ to generate, and we need
to pay attention to the number of samples that survive the rejection step, $S$.
We also may need to check that we have resolved the dominant modes enough that
$\mathcal{G}_{\rm max}$ is a good estimate of the true maximum of the marginal
likelihood.

We have implemented a few new features in \harv\ that help make the rejection sampling
more efficient and robust, or that differ from the default behavior of \joker, as we
describe below.

\subsubsection{Returning a fixed number of samples}

The number of samples that survive the rejection step, $S$, will depend on the data.
If a source has very informative data (many epochs, high signal-to-noise, etc.), then
$S$ will be small and sometimes even just one.
In the opposite regime (i.e. few epochs or low signal-to-noise data), $S$ can be very
large.
When processing many sources, it is more convenient (and, with JIT-compiled code, much
faster) to return the same number of samples for every source.
We therefore optionally replace the fourth step above.
Instead of rejecting samples, \harv\ can return the $S$ prior samples with the largest
marginal likelihood values for a fixed $S$, along with their importance weights
\begin{equation}
  w_j = \frac{\mathcal{G}_j}{\sum_{j'=1}^{J} \mathcal{G}_{j'}}
  \quad ,
  \label{eq:weights}
\end{equation}
which are normalized over all $J$ prior samples.
These samples are then weighted samples from the posterior pdf (not direct samples), but
they give us a diagnostic that we can use to assess the quality of the returned samples.
The sum of the $S$ importance weights (Equation~\ref{eq:weights}) is the fraction of the
posterior mass that the returned samples contain, so it should be close to one if $S$ is
large enough.
Another useful diagnostic is the effective sample size (ESS) of the weights,
\begin{equation}
  J_{\rm eff} = \frac{1}{\sum_{j=1}^{J} w_j^2}
  \quad ,
  \label{eq:ess}
\end{equation}
which is roughly the number of independent posterior samples that the $J$ prior samples
represent.

\subsubsection{Scalings with data quality}
\label{sec:methods-scaling}

The number of prior samples that the rejection sampler needs depends on how informative
the data are.
We can quantify this with the information gain, or just ``the information'' in the data
\citep{Skilling:2006},
\begin{equation}
  \mathcal{H} = \int \dd\vec{\theta}\, p(\vec{\theta} \given \vec{y})
    \ln \frac{p(\vec{\theta} \given \vec{y})}{p(\vec{\theta})}
  \quad ,
  \label{eq:information}
\end{equation}
which is the Kullback--Leibler divergence of the posterior pdf from the prior pdf over
the sampled parameters (the marginalized linear parameters do not contribute).
In this context, the posterior roughly occupies a fraction $e^{-\mathcal{H}}$ of the
prior mass \citep{Skilling:2006}, so each prior sample has a probability of about
$e^{-\mathcal{H}}$ of falling in the region of parameter space that has most of the
posterior mass.
The rejection sampler therefore returns $J_{\rm eff} \lesssim J\,e^{-\mathcal{H}}$
effective samples of the posterior (Equation~\ref{eq:ess}), and so the number of prior
samples it needs grows exponentially with $\mathcal{H}$.

Nested sampling \citep{Skilling:2004, Skilling:2006} also explores the full prior pdf,
but scales differently with $\mathcal{H}$ (see \citealt{Ashton:2022, Buchner:2023} for
reviews).
Nested sampling keeps a set of $n_{\rm live}$ ``live points,'' which are essentially
samples from the prior pdf that are constrained to have a likelihood above a threshold.
These points are repeatedly updated with new samples from the prior pdf that are
constrained to have en even higher likelihood than the lowest-likelihood live point.
Nested sampling keeps the replaced (``dead'') points and weights them by the likelihood
and the prior volume, so the dead points can be used to generate posterior samples and
to estimate the evidence.
Each live point that is replaced shrinks the prior volume enclosed by the live points by
a factor of about $e^{-1/n_{\rm live}}$, so every $n_{\rm live}$ replaced points shrink
the enclosed volume by another factor of $e$ (the constant, not the eccentricity).
To reach the posterior pdf, nested sampling therefore has to replace about
$n_{\rm live}\,\mathcal{H}$ points, so its cost grows only linearly with $\mathcal{H}$
\citep{Skilling:2006}.
The rejection sampler is efficient for uninformative data (small $\mathcal{H}$) but
becomes prohibitive for informative data, where nested sampling is the better choice.

\subsubsection{Validating the maximum marginal likelihood}
\label{sec:methods-gmax}

When only a few samples survive the rejection step, $\mathcal{G}_{\rm max}$ (the
threshold used in the rejection sampling) can still be much smaller than the height of
the true highest peak of the marginal likelihood.
Posterior samples that are returned in this regime may therefore not be representative
of the true posterior pdf, and may not even be in the dominant mode.
This commonly happens when the data are very informative, often with a long time
baseline, which leads to a highly structured posterior pdf with many narrow modes.
In many such instances, even with a large number of prior samples, the rejection
sampler can return one or just a few samples.
If one then initializes MCMC from these samples, the chains will usually stay in that
mode.
If that mode is not the dominant mode, the standard convergence diagnostics will not
reveal any issues, because all of the chains start in the same mode.
For example, the Gelman--Rubin statistic $\hat{R}$ \citep{Gelman:1992} compares the
variance between chains with the variance within each chain, and it is close to 1 when
the chains agree, even if they all sit in the wrong mode.

We recommend two diagnostic checks if the rejection sampling returns one or only a few
samples even when the number of prior samples $J$ is large.
First, rerun the rejection sampling with even more prior samples (e.g., $10J$): if
$\mathcal{G}_{\rm max}$ rises substantially, or the returned samples move to a different
mode, the first run had not located the dominant mode, and you should use the larger run
(or run with even more prior samples!), or switch to a different inference method (e.g.,
nested sampling).
Second, after continuing with MCMC, check the goodness of fit (e.g., $\chi^2$) at the
maximum a posteriori (MAP) sample.
With well-calibrated uncertainties, a large $\chi^2$ suggests that either the returned
sample(s) are not in the dominant mode or the model cannot explain the data.

Even if both of these checks pass, it still does not guarantee that we have
representative samples from the dominant posterior mode.
However, in practice, we find that unless the data are very adversarial, these
tests tend to be sufficient.
Section~\ref{sec:cases-apogee} shows an example of using these checks to assess the
returned samples.

\subsubsection{Evidence estimates and other sampling metadata}
\label{sec:methods-evidence}

The fully marginalized likelihood (FML) $p(\vec{y})$, or the evidence, is the integral
of the marginal likelihood over the prior pdf of the nonlinear parameters,
\begin{equation}
  p(\vec{y}) = \int \dd\vec{\theta} \,p(\vec{y} \given \vec{\theta})\,p(\vec{\theta})
  \quad .
\end{equation}
The FML is useful for model comparison, and we can compute it as a byproduct of the
rejection sampling procedure.
If we have samples $\vec{\theta}_j$ from the prior pdf, then we can estimate the
FML as
\begin{equation}
  p(\vec{y}) \approx \frac{1}{J} \sum_{j=1}^{J} \mathcal{G}_j
  \quad .
\end{equation}
This is a Monte Carlo estimate of the FML integral that uses all $J$ prior samples
(which are needed anyway for the rejection sampling procedure).
The estimate is unbiased, but it can have high variance, and the variance of the
estimate depends on the ESS $J_{\rm eff}$ (Equation~\ref{eq:ess}).
$J_{\rm eff}$ also tells us whether the sampling of the prior pdf is dense enough to
resolve the posterior pdf.
If $J_{\rm eff} \gg 1$, many prior samples contribute to the posterior pdf and the
uncertainty on $\ln p(\vec{y})$ is approximately $J_{\rm eff}^{-1/2}$.
If $J_{\rm eff} \approx 1$, a single prior sample dominates, and the posterior pdf is
probably much narrower than the typical spacing between prior samples.
In this regime, the returned samples may lie near a mode of the posterior but do not
necessarily resolve it.
Also, since the mode may not be the dominant mode of the posterior, the FML estimate
will be unreliable (but is typically underestimated).
On request, \harv\ returns these diagnostics and the FML estimate along with the
samples.

There are three possible outcomes of the rejection sampling for a given source that we
can use the ESS $J_{\rm eff}$ to assess:
\begin{enumerate}
  \item Many samples survive the rejection, $J_{\rm eff} \gg 1$. In this case, we are
    done for that source, and the FML estimate is likely reliable. This is the typical
    outcome for sparse or noisy data (Section~\ref{sec:cases-apogee}).
  \item Few samples survive, $J_{\rm eff} \sim 1$, and these samples are unimodal in
    period (i.e., their range in period is smaller than $\Delta = 4\,P_{\rm min}^2 /
    (2\pi\,T)$, where $P_{\rm min}$ is the smallest period among the samples and $T$ is
    the time spanned by the data, as in \citealt{Price-Whelan:2017}). In this case, the
    posterior pdf may be effectively unimodal, so we can use the surviving samples as an
    initialization for standard inference methods (e.g., MCMC), with the checks
    described in Section~\ref{sec:methods-gmax}. The FML estimate is then likely an
    underestimate of the true FML. This is the typical outcome for data
    with a high signal-to-noise ratio (Section~\ref{sec:cases-gaia4}).
  \item Few samples survive, $J_{\rm eff} > 1$ (but is small), and the samples span a
    period range larger than $\Delta$.
    In this case, the FML estimate is unreliable and probably also an underestimate of
    the true FML, but not always.
    This outcome can occur for data that are informative but with very non-uniform time
    sampling (with respect to the period of the companion), which can lead to a
    multimodal posterior pdf with narrow modes.
    In the context of \harv, these cases need even more prior samples, and this is
    expensive.
\end{enumerate}

\subsubsection{Using the returned samples in a hierarchical inference}

To use the returned samples within a hierarchical inference, one can use the importance
weights to reweight the samples to a population prior pdf, $p(\vec{\theta} \given
\vec{\alpha})$, that depends on the hyperparameters $\vec{\alpha}$ of a population
model.
For source $n$ with data $\vec{y}_n$, the likelihood given the hyperparameters is an
importance-sampling sum over the ratio of the population prior pdf to the interim prior
pdf, $p(\vec{\theta} \given \vec{\alpha}_0)$, that was used for the sampling
\citep{Hogg:2010},
\begin{align}
  p(\vec{y}_n \given \vec{\alpha}) &\approx
    Z_{{\rm int},n} \sum_{k=1}^{S_n} w_{nk}\, r_{nk}(\vec{\alpha})
  \label{eq:reweight} \\
  r_{nk}(\vec{\alpha}) &=
    \frac{p(\vec{\theta}_{nk} \given \vec{\alpha})}
      {p(\vec{\theta}_{nk} \given \vec{\alpha}_0)}
  \quad ,
\end{align}
where $Z_{{\rm int},n}$ is the evidence for $\vec{y}_n$ under the interim prior, the sum
is over the $S_n$ returned samples $\vec{\theta}_{nk}$, and $w_{nk}$ are their weights
(Equation~\ref{eq:weights}).
If we instead use the rejection sampler to return posterior samples, then $w_{nk} =
1/S_n$ and the sum is the mean of the prior ratio over the posterior samples.
For a population of sources, the likelihood is the product of Equation~\ref{eq:reweight}
over the sources.
If every source uses the same interim prior and the same model, the evidence terms
$Z_{{\rm int},n}$ do not depend on $\vec{\alpha}$ and can be dropped (but see
Section~\ref{sec:cases-pop-inf} for a case where we must compute and use them).

\subsection{Using \harv\ models with other tools}
\label{sec:methods-interop}

For a given prior pdf and data set, \harv\ can construct a \package{numpyro} model such
that the parameters are sampled from the prior pdf and the marginal likelihood
(Equation~\ref{eq:marginal-likelihood}) enters as a factor in the model.
This is the model that the MCMC sampler in \harv\ runs (\texttt{NumpyroSampler} in the
code).
This can also be used with any other inference method in \package{numpyro} (e.g.,
variational inference), or even with other libraries that can use \jax\ functions.
For example, \package{numpyro} enables extracting a function that evaluates the log
density (and components) of a model, which can be passed as \jax\ functions to samplers
in other libraries.
We use this scheme in Section~\ref{sec:cases-interop} to run a nested sampler from
\package{blackjax} \citep{Cabezas:2024} on some of the sources from
Section~\ref{sec:cases-apogee}.

We can also use the marginal log-likelihood function itself for data formats that \harv\
supports within a larger or separate model to handle additional data or parameters.
For example, imagine that we have RV and light curve data for a system where the
companion also transits the host star.
The transit light curve depends on period, eccentricity, argument of pericenter, and
phase (as do the RVs), along with parameters that do not appear directly in the RV model
(the inclination or impact parameter, the radius ratio, and the limb darkening).
In this case, for example, we could use \package{jaxoplanet} \citep{jaxoplanet} to
compute the transit light curve and its log likelihood, and then use \harv\ to compute
the marginal log likelihood of the RVs --- the full log likelihood function would still
be a JIT-able \jax\ function that can then be used with other inference methods.
Or, we could construct a joint model in \package{numpyro} that samples the shared
orbital parameters and the additional transit parameters, and adds the log likelihoods
of the two data sets.
We do not demonstrate this here, but mention it only as a possibility given the
flexibility of tools built on \jax\ and \package{numpyro}.

\section{Case studies}
\label{sec:case-studies}

In this section we demonstrate \harv\ with four case studies.
We have chosen these case studies to illustrate different aspects of the package and
to highlight some best pracices.
These are demonstrations of the methods and we do not consider the results for the
individual systems to be new astrophysical results.
The first three case studies use published data, and the last case study uses simulated
data so that we know the true population parameters.
Each case study is reproduced by a set of scripts or Jupyter notebooks in the repository
that accompanies this article,\footnote{See the ``scripts'' directory in
\url{https://github.com/adrn/harv-paper}.} using version \harvversion\ of \harv.

\subsection{Case study 1: APOGEE radial velocities for binary-star systems}
\label{sec:cases-apogee}

\begin{table*}
  \centering
  \caption{
    Results for the six \apogee\ sources, using the same $J = 10^7$ prior
    samples, except for 2M15035796-4533457 and 2M20183197+1953430, for which we use a
    second run with $J = 10^8$ new prior samples (see text).
    $N$ is the number of visits, $T$ is the time spanned by the visits, $S$ is the
    number of samples that survive the rejection step, $J_{\rm eff}$ is the effective
    sample size, and $\ln\mathcal{B}$ is the log Bayes factor of the companion model
    relative to a constant-velocity model.
    For sources with $J_{\rm eff} < 3$, the evidence estimate and therefore
    $\ln\mathcal{B}$ is unreliable, so we mark these values with a dagger.
    \label{tab:apogee}
  }
  \begin{tabular}{lrrrrrl}
    \hline
                 &        &          & Number of           & Effective     & Log Bayes        & Sampling \\
     \apogee\ ID & Visits & Baseline & Rejection samples   & sample size   & factor           & outcome  \\
                 & $N$    & $T$ [\unit{day}] & $S$                 & $J_{\rm eff}$ & $\ln\mathcal{B}$ &          \\
    \hline
    2M15323868+4218297 & 3 & 26 & \num{9138} & \num{1.84e+04} & $20401$ & rejection \\
2M02323271+3350542 & 3 & 59 & \num{1353023} & \num{2.75e+06} & $2$ & rejection \\
2M15035796-4533457 & 6 & 1005 & \num{2} & \num{2.04} & $55354^\dagger$ & needs more prior samples \\
2M22122067+4152581 & 6 & 681 & \num{666393} & \num{1.25e+06} & $3$ & rejection \\
2M20183197+1953430 & 13 & 761 & \num{1} & \num{1} & $97554^\dagger$ & MCMC \\
2M16022747+2902099 & 12 & 2524 & \num{40352} & \num{2.06e+05} & $22$ & rejection \\

    \\
    \hline
  \end{tabular}
\end{table*}

This first case study demonstrates the basic functionality of \harv\ and shows the
typical pipeline for analyzing a small sample of stars that may have varied data
quality and time sampling.
For this example, we use visit-level radial velocity (RV) measurements from \apogee\
\dr{17} \citep{Majewski:2017, SDSSDR17} for a few handpicked sources that have a
range of numbers of visits.
We select six sources: two with 3 visits, two with 6 visits, and two with around 12
visits (Table~\ref{tab:apogee}).
For each pair of sources (per number of visits), one has large RV scatter (an obvious
binary) and one has small RV scatter (unclear).

We use the standard RV parameterization and a prior pdf that is similar to the prior we
used previously to analyze \apogee\ data with \joker\ \citep{Price-Whelan:2020}.
This prior is uniform in $\ln P$ between 2 and $16{,}384~\unit{day}$, a truncated normal
distribution for the eccentricity with mean 0.5 and standard deviation 0.25 (truncated
to $[0, 1)$), and uniform distributions for the angles and phases $\omega$ and $\tau$.
For the linear parameters, we use the period-dependent prior scale for the velocity
semi-amplitude $K$ (Equation~\ref{eq:sigma-K}) with $\sigma_{K,0} =
30~\unit{km.s^{-1}}$, $P_0 = 1~\unit{\year}$, and a normal prior for the systemic
velocity $v_0$ with standard deviation $\sigma_{v_0} = 150~\unit{km.s^{-1}}$.
We do not include a jitter parameter here, and we discuss the visit velocity
uncertainties below.

\begin{figure*}
  \centering
  \includegraphics[width=\textwidth]{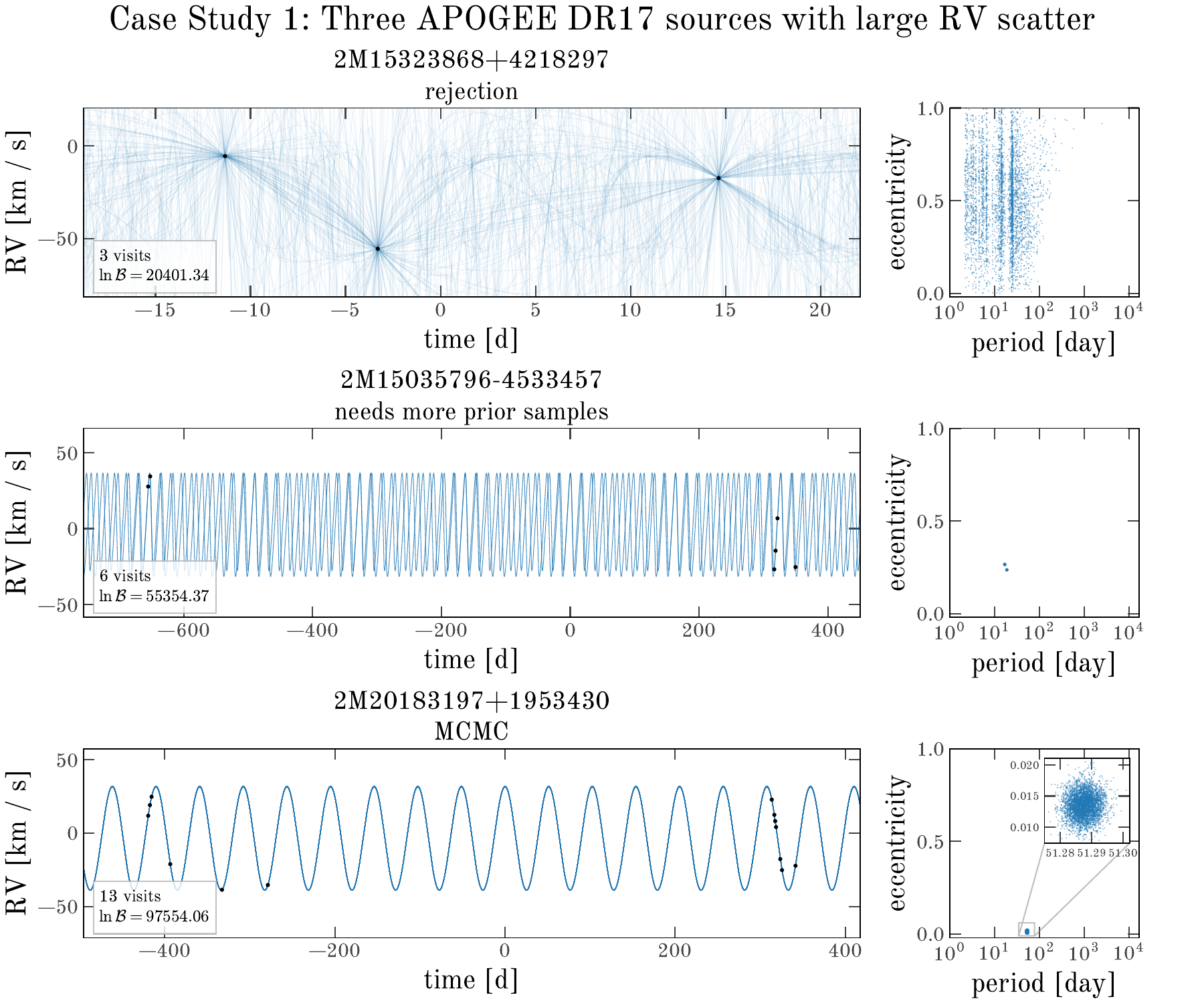}
  \caption{
    Data and posterior samples for the three \apogee\ sources with large RV scatter,
    ordered by the number of visits (top to bottom).
    For each source (row), the left panel shows the visit velocities (black points) with
    orbits computed from up to 128 posterior samples (blue lines), and the right panel
    shows a larger set of posterior samples in period and eccentricity.
    The inset box in the left panel shows the number of visits and the log Bayes factor
    $\ln\mathcal{B}$ for the companion model relative to a constant-velocity model.
    The samples are from the rejection sampling (with $10^8$ prior samples for
    2M15035796-4533457 and 2M20183197+1953430), except for 2M20183197+1953430, where we
    also continue sampling with MCMC.
    \label{fig:apogee-large}
  }
\end{figure*}

\begin{figure*}
  \centering
  \includegraphics[width=\textwidth]{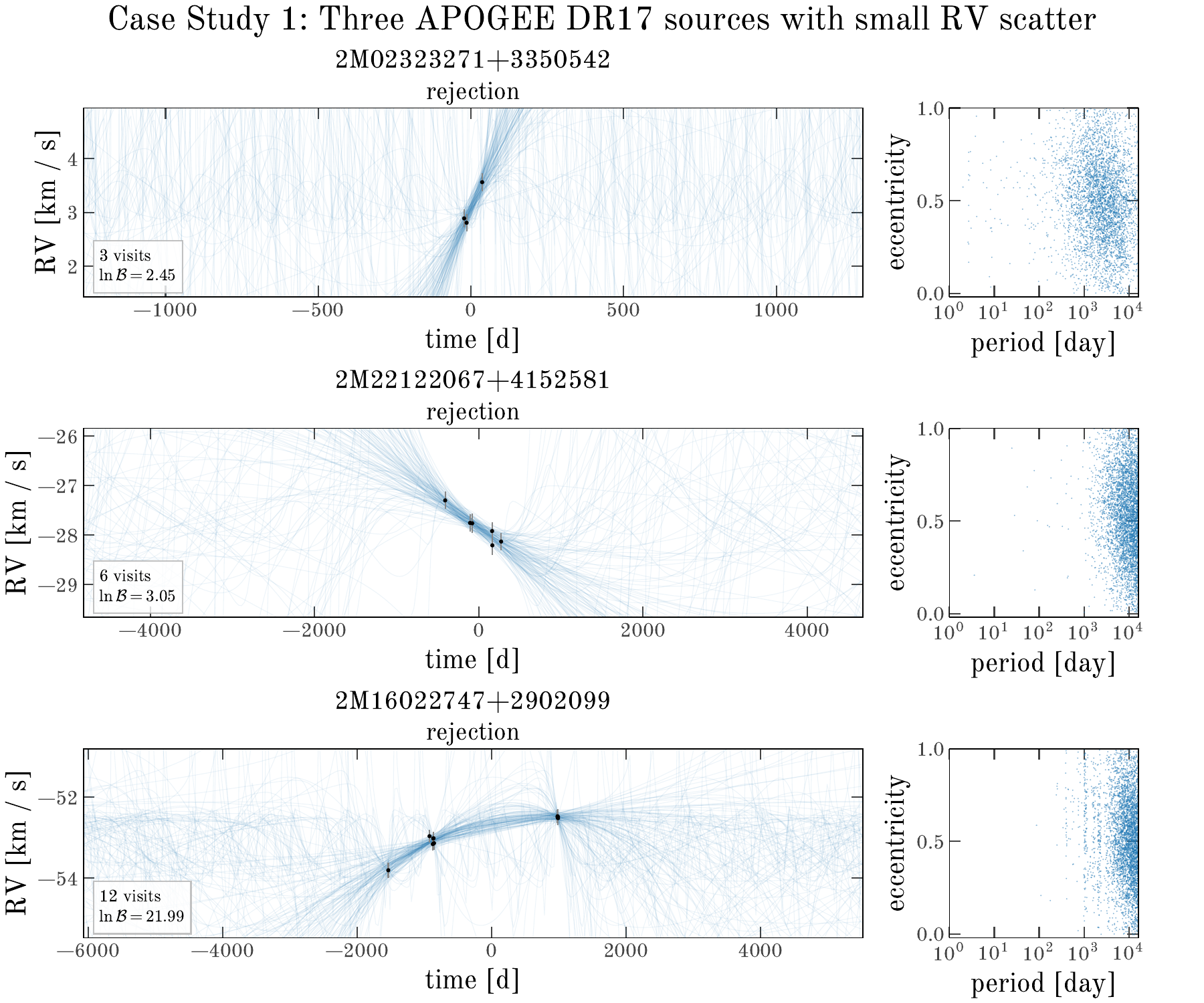}
  \caption{
    Same as Figure~\ref{fig:apogee-large}, but for the three sources with small RV
    scatter. All samples are from the rejection sampling.
    \label{fig:apogee-small}
  }
\end{figure*}

We first generate $J = 10^7$ prior samples, and then use the same prior samples for all
stars.
For each source, we run the rejection sampling with these prior samples, and request the
evidence estimate and the diagnostics described in Section~\ref{sec:methods-rejection}:
\begin{lstlisting}
samples = sampler.run_with_samples(
    data, prior_samples, key=key,
    return_evidence_stats=True)
\end{lstlisting}
\noindent
Figures~\ref{fig:apogee-large} and~\ref{fig:apogee-small} show the data and the returned
samples for the six sources, and Table~\ref{tab:apogee} summarizes the results from this
analysis.
All three outcomes of the rejection sampling (Section~\ref{sec:methods-rejection})
appear among these six sources.
For the three sources with small RV scatter, many samples survive in all cases.
The samples cover most of the prior at long periods (mostly $P \gtrsim 100~\unit{day}$
with 3 visits and $\gtrsim 1000~\unit{day}$ with 6, but that is likely more a statement
about the time baseline of the data in each case).
For the three sources with large RV scatter, the final outcome depends on the number of
visits: For the source with 3 visits, many samples survive and nearly all have periods
$\lesssim 100~\unit{day}$, but they are spread over many period modes.
For the sources with 6 and 13 visits, only two samples and one sample survive,
respectively, so we rerun the rejection sampling for these two sources with $10^8$ new
prior samples (Section~\ref{sec:methods-gmax}), and we show and report the results of
these larger runs (see below).

For the two sources with $N=3$ visits, the rejection sampler returns many samples with a
large effective sample size; the data are fairly uninformative, and the posterior pdf is
broad and multimodal.
However, note that the source with large scatter (2M15323868+4218297) has a much larger
Bayes factor than the source with small scatter (2M02323271+3350542), which suggests
that the data prefer the companion model for the former, while the preference is
marginal for the latter.

For the two sources with $N=6$ visits, the rejection sampler returns many samples for
the one with small scatter (2M22122067+4152581), which also has a large effective sample
size and a marginal Bayes factor.
For the source with large scatter (2M15035796-4533457), only two of the $10^7$ prior
samples survive, in different period modes, so we rerun the rejection sampler with
$10^8$ new prior samples.
With $10^7$ prior samples, the best sample has $\ln \mathcal{G}_{\rm max} = -21.4$ and
the two samples have periods around $33~\unit{day}$.
With $10^8$ prior samples, the best sample has $\ln \mathcal{G}_{\rm max} = -11.9$ and,
again, two samples survive, but now with periods around $17~\unit{day}$.
Even with $10^8$ prior samples, the effective sample size is small, and the samples are
still sparse, which means that we need even more prior samples to resolve the modes.
The evidence estimate is also unreliable, but the Bayes factor is so large
($\ln\mathcal{B} \approx 5.5\times10^4$) that the data strongly prefer the companion
model, even allowing for a large error in the estimate.
For cases like this, the simplest path forward is to rerun the rejection sampler with
\textit{even more} prior samples: the ratio $J_{\rm eff}/J \approx 2\times10^{-8}$
implies that we would need of order $10^9$--$10^{10}$ prior samples to return a few
hundred posterior samples.
That is expensive, but possible for a small number of sources, and it is straightforward
to parallelize this calculation (or to run it on a GPU for better performance;
Section~\ref{sec:performance}).

For the two sources with $N \approx 12$ visits, the rejection sampler returns many
samples for the source with small scatter (2M16022747+2902099), with a large effective
sample size, but the Bayes factor is moderately large, suggesting that the data
prefer the companion model even if the orbital parameters are not well constrained.
For the source with large scatter (2M20183197+1953430), only one of the $10^7$ prior
samples survives, at $P = 55.24~\unit{day}$ with $\ln \mathcal{G}_{\rm max} = -312.4$,
so we rerun the rejection sampling with $10^8$ new prior samples:
\begin{lstlisting}
samples = sampler.run(data, key=key,
    n_prior_samples=100_000_000,
    return_evidence_stats=True)
\end{lstlisting}
\noindent
With $10^8$ prior samples, the best sample has $\ln \mathcal{G}_{\rm max} = -115.7$ and
a period $P = 51.29~\unit{day}$.
At this point, still only one sample survives, so we continue sampling with MCMC.
For this case, we initialize the No-U-Turn Sampler \citep[NUTS;][]{Hoffman:2011}
implemented in \package{numpyro} using the single rejection sample:
\begin{lstlisting}
if samples.period_unimodal(data):
    mcmc = harv.NumpyroSampler(prior, model)
    samples = mcmc.run(
        data, init_samples=samples, key=key)
\end{lstlisting}
\noindent
We run 4 chains, each with 1000 warm-up steps followed by 2000 sampling steps.
At the end of the sampling, all parameters have $\hat{R} < 1.01$.
At the MAP sample, $P = 51.29~\unit{day}$ and $\chi^2 \approx 3$ for 13 visits (7
degrees of freedom), which is a good fit given the calibrated uncertainties.
Note that if we had continued with MCMC from the single sample from the first run with
$10^7$ prior samples, all chains would have converged within a peak at $P =
55.21~\unit{day}$ (also with $\hat{R} < 1.01$ for all parameters, because all of the
chains start in the same mode), but with $\chi^2 = 418$ at the MAP sample.
This emphasizes the importance of checking the goodness of fit at the MAP sample, and
rerunning the rejection sampling with more prior samples if the fit is bad.
However, a good fit still does not rule out that other (better) modes exist in the
posterior pdf, and unfortunately the only way to know that is to continue running the
rejection sampling with more and more prior samples until the returned samples are dense
enough to resolve the posterior structure.
For this case, we compare our results with those from a nested sampler, which explores
the full prior, in Section~\ref{sec:cases-interop}.

Because we do not include a jitter parameter or other noise model, both the companion
model and the constant-velocity model (used for the evidence and Bayes factor estimates)
assume that any scatter in the velocities that cannot be explained by the
constant-velocity model comes from a companion.
The \apogee\ catalog uncertainties are known to be underestimated
\citep[e.g.,][]{Badenes:2018, Price-Whelan:2020}, so for all of the results above we use
the recalibrated visit uncertainties from \citet{Price-Whelan:2020}.
If we instead use the raw \apogee\ catalog uncertainties, the log Bayes factor for the
marginal sources is substantially higher.
Instead of recalibrating the uncertainties, another way to handle this would be to add
a jitter parameter for each source and marginalize over its value.
We do not recommend this for large samples because it will limit the detection
efficiency of low-amplitude companions, and it makes the rejection sampling less
efficient (it adds another nonlinear parameter that must be sampled explicitly).
A better approach is to first calibrate the uncertainties with an independent
model fit to the catalog as a whole, and then use the calibrated uncertainties for the
per-source inferences, as we have done in \citet{Price-Whelan:2020}.

\subsection{Case study 2: Modeling Gaia-4 with radial velocities and DR4 epoch
astrometry}
\label{sec:cases-gaia4}

\begin{figure*}
  \centering
  \includegraphics[width=\textwidth]{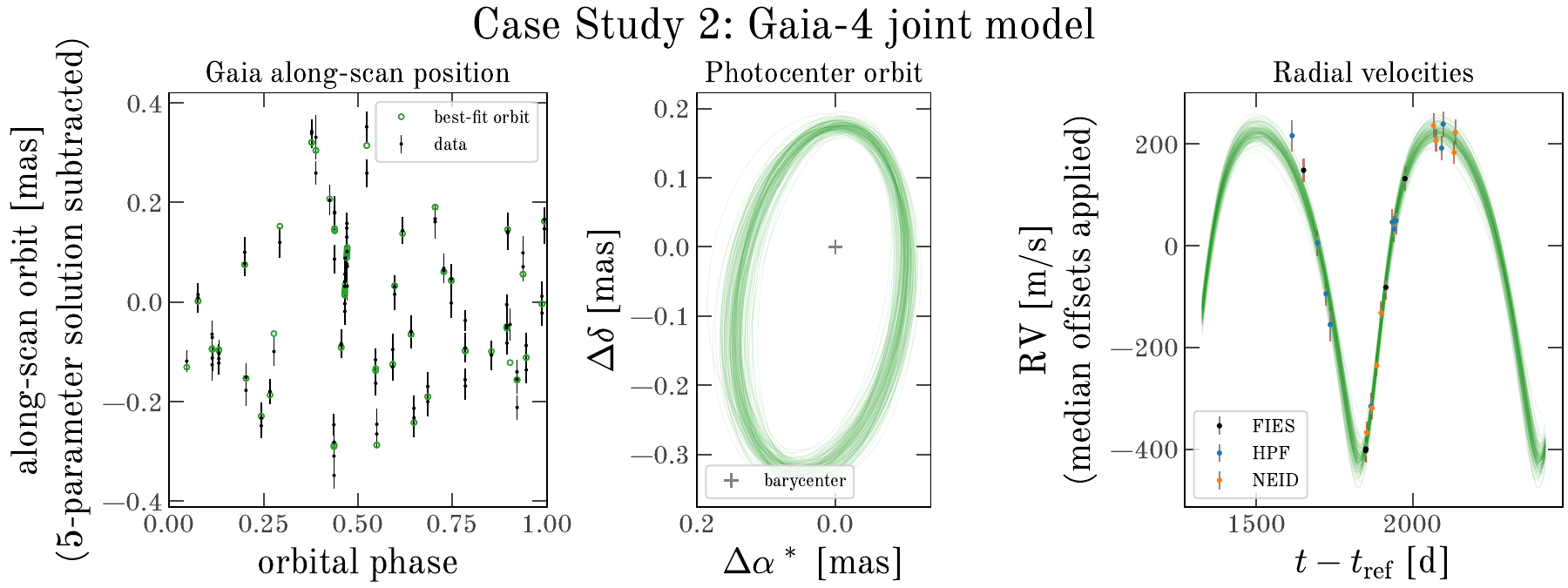}
  \caption{
    Results from the joint model for Gaia-4.
    \textit{Left:} The \gaia\ \dr{4} along-scan positions after subtracting the
    five-parameter (sky position, proper motion, and parallax) astrometric solution of
    the best-fit model. Black points show the data as a function of orbital phase, and
    green open markers show the model predictions at the same epochs.
    \textit{Middle:} The orbit of the photocenter on the sky for 128 posterior samples.
    \textit{Right:} The RVs from the three instruments, corrected with the inferred
    (median) offset values, and the model predictions for 128 posterior samples.
    \label{fig:gaia4}
  }
\end{figure*}

\begin{table*}
  \centering
  \caption{
    Posterior constraints for Gaia-4b from the astrometry alone and from the joint
    model (median and 68\% interval), compared with published values.
    The companion mass assumes a stellar mass of $0.644~\unit{\Msun}$
    \citep{Stefansson:2025}.
    \label{tab:gaia4}
  }
  \begin{tabular}{lcccc}
    \hline
    Parameter & Astrometry only & Joint model & \citet{Kanodia:2026} &
      \citet{Stefansson:2025} \\
    \hline
    $P$ [\unit{day}] & $577.6 \pm 3.7$ & $575.6 \pm 1.1$ & $575.90 \pm 0.90$ & $571.3 \pm 1.4$ \\
$e$ & $0.386 \pm 0.077$ & $0.324 \pm 0.020$ & $0.319 \pm 0.012$ & $0.338^{+0.026}_{-0.023}$ \\
$\cos i$ & $-0.519 \pm 0.043$ & $-0.500 \pm 0.031$ & $-0.508 \pm 0.024$ & $-0.452 \pm 0.067$ \\
$a_0$ [\unit{mas}] & $0.252 \pm 0.011$ & $0.2534 \pm 0.0061$ & $0.2500 \pm 0.0040$ & $0.279^{+0.016}_{-0.014}$ \\
$\varpi$ [\unit{mas}] & $13.6246 \pm 0.0081$ & $13.6254 \pm 0.0079$ & $13.6251 \pm 0.0074$ & \nodata \\
$K$ [\unit{m.s^{-1}}] & \nodata & $321.9 \pm 9.9$ & $315.1 \pm 3.8$ & $368^{+22}_{-18}$ \\
$M_{\rm p}$ [\unit{\Mjup}] & $10.77 \pm 0.50$ & $10.84 \pm 0.35$ & $10.70 \pm 0.32$ & $11.80^{+0.73}_{-0.66}$ \\

    \\
    \hline
  \end{tabular}
\end{table*}

This second case study demonstrates how to use \harv\ to model a source with both radial
velocity and astrometric data, and with very informative data that precisely constrain
the orbital parameters.
Gaia-4b is a super-Jupiter planet with a $P\approx 570~\unit{day}$ orbit around a nearby
K dwarf star.
The planet was initially identified as a candidate from its astrometric orbit in \gaia\
\dr{3} and it was confirmed with RVs \citep{Stefansson:2025}.
This source was included in the \gaia\ \dr{4} pre-release, so we have access to the
epoch astrometry data as a representative example of the data that will be available in
\gaia\ \dr{4}.
From the \gaia\ \dr{4} pre-release, we have 93 field-of-view transit measurements of
this source, which we combine with 23 RVs from three spectrographs (10 from HPF, 8 from
NEID, and 5 from FIES; \citealt{Stefansson:2025}) to jointly model the properties of the
planet.
While we were preparing this manuscript, \citet{Kanodia:2026} published a joint analysis
of the Gaia-4 system using the same data, so we compare our results to theirs below (but
they use additional RV observations).

The \gaia\ \dr{4} pre-release (and \gaia\ \dr{4} main data) provides the epoch
astrometry for each CCD within each field-of-view transit.
Given the rotation rate of the \gaia\ spacecraft, it takes $\lesssim 60$ seconds for a
source to cross the full astrometric field (all nine CCDs).
Since this time is much shorter than the orbital periods of companions we consider, we
combine the CCD measurements into a single ``transit-level'' measurement per
field-of-view crossing.
We only keep the CCD measurements that have the \texttt{used\_by\_agis\_al} flag set to
true, and take the inverse-variance weighted mean of the along-scan positions of the
per-CCD measurements.
We keep the mean observation time and scan angle for each transit (the parallax factor
is already given per transit).
Of the 115 transits of Gaia-4, only 93 have CCD measurements that pass the above cut.
The median astrometric uncertainty of the remaining transits is $\approx
0.028~\unit{mas}$ per transit.
We do not add the excess noise of the source to these uncertainties, because that might
be inflated by the influence of the planet itself (instead, we fit for an astrometric
jitter term; see below).

We start modeling the data using only the astrometry with the rejection sampler.
We use the standard parameterization for the astrometric model, with a prior pdf that
is uniform in $\ln P$ between 10 and $2000~\unit{day}$ and the default prior pdfs for
the other nonlinear parameters.
We adopt Gaussian prior pdfs for the linear parameters in which the reference position
has a scale of $100~\unit{mas}$, and the proper motion components and the angular
semi-major axis of the photocenter orbit have parallax-dependent scales, set by a
tangential velocity scale of $100~\unit{km.s^{-1}}$ and by $\sigma_{a,0} = 1~\unit{au}$
(Equation~\ref{eq:sigma-a}), respectively.
We compute these scales from the \gaia\ \dr{3} parallax, $13.643~\unit{mas}$, so that
they do not depend on the parallax parameter, and the parallax is marginalized along
with the other linear parameters (Section~\ref{sec:methods-priors-nonlinear}).
This assumes that the orbit does not bias the catalog parallax much, which holds here
because the photocenter orbit is small compared with the parallax and the period is not
close to 1 year.
We also use the catalog parallax to set the prior on the parallax: a normal distribution
centered on $13.643~\unit{mas}$ with a width of $1~\unit{mas}$, which is broad compared
to the size of the orbit and the catalog parallax uncertainty.
We also include an astrometric jitter parameter, with a half-normal prior pdf with a
scale of $0.1~\unit{mas}$, because the \gaia\ \dr{4} pre-release data are not yet fully
calibrated and may have underestimated uncertainties.
With these choices, we generate $J = 10^7$ prior samples and run the rejection sampling.
Only one sample survives the rejection step, with a period of $P \approx
574~\unit{day}$.

We continue with MCMC again using NUTS.
We first use the returned sample to initialize an optimizer to find the nearby maximum
of the posterior pdf mode that the sample is in, which shifts the MCMC initialization to
$P = 578.3~\unit{day}$, and the goodness of fit here is (reduced) $\chi^2 = 1.02$ (see
Section~\ref{sec:methods-gmax}).
We then run 4 NUTS chains from this location, each with 1000 warm-up steps followed by
2000 sampling steps.
From modeling the astrometry alone, we infer $P = 577.7 \pm 3.7~\unit{day}$ and $e =
0.39 \pm 0.08$ (Table~\ref{tab:gaia4}).

We then incorporate RV data into the model.
The joint model shares the orbital parameters between the two data types (RV and
astrometry).
Because the companion is unseen, the astrometric orbit determines the velocity
semi-amplitude:
\begin{lstlisting}
joint_model = harv.JointModel.for_rv_and_gaia(
    components={"rv": rv_model,
                "astrom": astrom_model})
\end{lstlisting}
\noindent
In the joint model, \harv\ samples the parallax explicitly because this is required to
relate the physical size of the astrometric orbit to the velocity semi-amplitude.
The RV model here includes two new parameters that represent the velocity offset for HPF
and NEID relative to FIES, which are linear parameters that \harv\ marginalizes over,
along with a jitter parameter shared by all three instruments:
\begin{lstlisting}
rv, indicator, names = rv_data.indicator_data_by_type(
    harv.RVData, reference="FIES")
rv_model = harv.RVModel(extensions=(
    harv.Jitter(obs_unit="m/s"),
    harv.MultiSurveyOffset(indicator, names, "m/s")))
\end{lstlisting}
\noindent
The astrometry alone cannot distinguish the orbit from a mirrored orbit in which
($\omega \to \omega + \pi$, $\Omega \to \Omega + \pi$), so we first optimize the joint
posterior pdf starting from both, and initialize the MCMC at the one with the larger
posterior probability.
The two optima have a difference of $\Delta \ln p \approx 82$, so the RV data clearly
break this degeneracy.

Figure~\ref{fig:gaia4} shows the data and some predictions from the joint model.
Table~\ref{tab:gaia4} summaries the results and compares our inferred parameter values
with previous work.
We infer offsets of HPF and NEID relative to FIES of $-115 \pm 15$ and
$-99 \pm 15~\unit{m.s^{-1}}$, respectively, which are consistent in magnitude with the
$\approx 100~\unit{m.s^{-1}}$ offsets found by \citet{Kanodia:2026} for these
instruments.
We infer a planet mass of $10.84 \pm 0.35~\unit{\Mjup}$, which is also consistent with
the recent analysis from \citet{Kanodia:2026}, $10.70 \pm 0.32~\unit{\Mjup}$, who use
the same \gaia\ pre-release data but more RV measurements.

\subsection{Case study 3: Connecting \harv\ with other inference tools and a comparison
to nested sampling}
\label{sec:cases-interop}

Because \harv\ builds on top of \package{numpyro} (Section~\ref{sec:methods-interop}),
which supports extracting functions that evaluate the log density components of a model,
we can use the same model in other inference tools, outside of \harv\ or
\package{numpyro} itself.
In this case study, we run a nested sampler \citep{Skilling:2004, Skilling:2006} with a
\harv\ model for four of the \apogee\ sources from Section~\ref{sec:cases-apogee}, and
compare the results with those from the rejection sampler.
Nested sampling explores the full prior pdf and returns posterior samples and the
evidence, so it is often used for posterior pdfs with many modes.
We therefore also use the comparison here to show why we do not use it in place of the
rejection sampler for every source.
We use the nested slice sampler \citep{Yallup:2026} implemented in \package{blackjax}
\citep{Cabezas:2024}.

The nested sampler needs the log prior pdf and the log likelihood as separate functions.
\package{numpyro} provides both of these from the model that is constructed in \harv, as
we show below.
In the code, the marginal likelihood (Equation~\ref{eq:marginal-likelihood}) enters the
model as a factor named \texttt{ln\_lik}:
\begin{lstlisting}
numpyro_model = model.numpyro_model(
    prior.nonlinear_priors, data,
    prior.linear_priors)
prior_model = block(numpyro_model,
                    hide=["ln_lik"])

def log_prior(z):
    return -potential_energy(
        prior_model, (), {}, z)

def log_lik(z):
    p = constrain_fn(prior_model, (), {}, z)
    return log_likelihood(numpyro_model, p,
        batch_ndims=0)["ln_lik"]

nss = blackjax.nss(log_prior, log_lik,
    num_inner_steps=20, num_delete=100)
\end{lstlisting}
\noindent
Here, \texttt{z} is a point in the unconstrained space used internally by
\package{numpyro} --- \package{numpyro} maps each parameter to the real line, and the
log prior pdf includes the Jacobian of this mapping.
For the nested sampling run, we draw 1000 live points from the prior pdf, take 20 slice
steps per new point, replace 100 points per iteration, stop when the evidence in the
live points is less than $10^{-3}$ of the accumulated evidence, and run the nested
sampler with five different seeds for each source.
We take the number of slice steps as 5 times the number of sampled parameters, which is
the recommended default in \package{PolyChord} \citep{Handley:2015}.
We do not explore the nested sampling settings exhaustively because the point here is
mainly to show how to use \harv\ with other inference tools and to compare the results
with those from the rejection sampler.

For the three sources from Section~\ref{sec:cases-apogee} that have large effective
sample sizes $J_{\rm eff} \gg 1$ --- 2M02323271+3350542, 2M16022747+2902099, and
2M15323868+4218297 --- the evidence estimates from the two methods agree.
With the rejection sampler, we get $\ln p(\vec{y}) = -8.4890 \pm 0.0005$, $-6.2214 \pm
0.0022$, and $-17.729 \pm 0.007$, and with nested sampling we get $-8.50 \pm 0.03$,
$-6.24 \pm 0.09$, and $-17.74 \pm 0.07$ (mean and standard deviation over random number
seeds).
For these sources, the uncertainty that each nested sampling run reports also matches
the scatter over the random seeds (0.034, 0.060, and 0.082 versus 0.031, 0.086,
and 0.069), which suggests that our adopted settings are sufficient for estimating the
evidence for these sources.

Nested sampling and \harv's rejection sampler are similar in spirit.
Both can estimate the evidence using samples from the prior pdf, and both handle
multimodal posterior pdfs better than traditional MCMC.
But their cost depends differently on the information $\mathcal{H}$ in the data: the
number of prior samples that the rejection sampler needs grows like $e^{\mathcal{H}}$,
whereas the number of nested sampling iterations grows like $\mathcal{H}$
(Section~\ref{sec:methods-scaling}).
The example data here confirm the scaling of the rejection sampler:
For the three sources above, the nested sampling runs give $\mathcal{H} = 1.05$, 3.27,
and 6.11, so $e^{-\mathcal{H}} = 0.35$, 0.038, and 0.0022, and the rejection runs from
Section~\ref{sec:cases-apogee} give $J_{\rm eff}/J = 0.28$, 0.021, and 0.0018 ---
slightly below these upper bounds, as expected (Section~\ref{sec:methods-scaling}), and
comparable in magnitude.

The nested sampling runs for 2M20183197+1953430 ($\mathcal{H} \approx 26$) take about
330 iterations, compared with about 130 for 2M15323868+4218297 ($\mathcal{H} \approx
6.1$), whereas the rejection sampler would need about $e^{20} \approx 5 \times 10^{8}$
times more prior samples, or about $J \gtrsim e^{26} \approx 10^{11}$ total prior
samples.
As we see above, even with $10^8$ prior samples, only one sample survives rejection, and
the evidence estimate ($-134.1$) is way below the nested sampling estimate ($-33.0 \pm
2.1$).
All five nested sampling runs find a single period mode at $P = 51.29~\unit{day}$, the
same mode as the MCMC samples in Section~\ref{sec:cases-apogee}.
For sources like this, nested sampling (or standard MCMC if the samples lie in a single
period mode) is a better option than the rejection sampler.
We still recommend running the rejection sampler first, to identify cases like this, and
then following up with nested sampling or MCMC.

\begin{figure*}
  \centering
  \includegraphics[width=\textwidth]{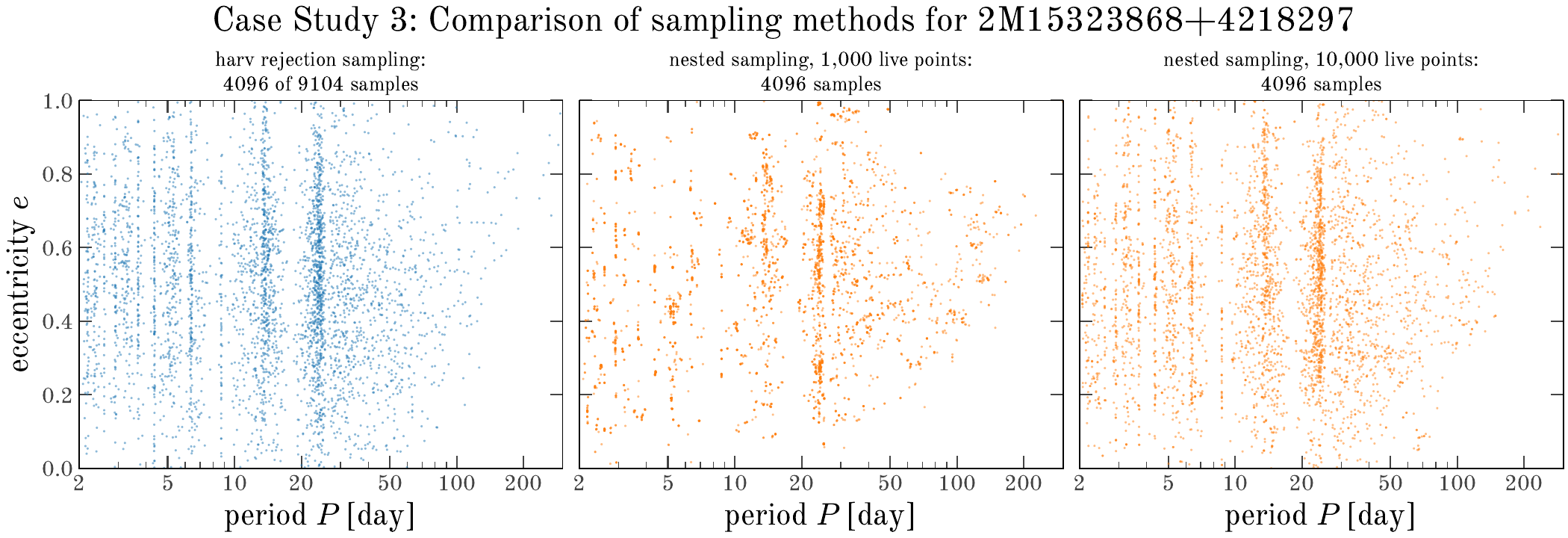}
  \caption{
    Posterior samples in period and eccentricity for 2M15323868+4218297 (see
    Figure~\ref{fig:apogee-large}, top left panel), which has three visits over about
    26~days.
    \textit{Left:} 4096 random samples (of 9104) from one run of \harv's rejection
    sampler with $10^7$ prior samples.
    \textit{Middle:} 4096 samples from one run of the nested slice sampler with 1000
    live points, drawn from the weighted dead points of the run (see text).
    \textit{Right:} The same, for a run with 10,000 live points.
    All of these runs give the same evidence and find the same period modes.
    With 1000 live points, the samples from nested sampling appear clumpy within the
    modes. This is likely because consecutive points are highly correlated and we do not
    have enough live points.
    With 10,000 live points, the samples look much more like the rejection samples, but
    each run takes about 20 times as long as a rejection run, and the probabilities of
    some modes differ between runs (see text in Section~\ref{sec:cases-interop}).
    \label{fig:nested-sampling}
  }
\end{figure*}

However, nested sampling, like MCMC, has its own limitations.
For example, nested sampling can return a bad sampling from a posterior pdf with many
modes even when the data are only moderately informative.
We demonstrate this with the source 2M15323868+4218297, which has three visits over 26
days and a large RV scatter (Figure~\ref{fig:apogee-large}).
From the rejection sampler, we can see that the posterior pdf has many modes in period
between 2 and about $100~\unit{day}$ (also see left panel of
Figure~\ref{fig:nested-sampling}).
For this source, we estimate $\mathcal{H} \approx 6.1$ from the rejection sampling run,
and with an initial sample of $J = 10^7$ prior samples, we get an ESS of $J_{\rm eff} =
1.8 \times 10^4$.
Here, we perform five runs of the rejection sampler with independent prior samples
and find that they agree to within the sampling noise.

We then run the nested slice sampler (with the same setup as above using 1000 live
points) five times with different random seeds.
We get a consistent evidence estimate from the nested sampler (compared to the rejection
sampler) and similar mode structure in period (at least visually; middle panel of
Figure~\ref{fig:nested-sampling}), but the posterior samples are much clumpier than the
rejection samples.
The 4096 samples in the middle panel are resampled from the weighted dead points, but
consecutive dead points along a slice walk are correlated, which can lead to poor
sampling of the posterior distribution.
Also, the probabilities of the modes vary between our five nested sampling runs by more
than the uncertainty implied by the effective sample size of each run (about 2200).
For example, the fraction of the posterior probability between 12.5 and
$15.5~\unit{day}$ (visibly, one of the dominant modes in
Figure~\ref{fig:nested-sampling}) ranges from 0.110 to 0.174 over the five nested
sampling runs, but the estimated uncertainty from a single run is only $\approx 0.007$.
If we repeat the rejection sampling with five random number seeds, the probability of
the same mode varies from 0.127 to 0.132, consistent with the estimated uncertainty of
0.004.
Importantly, running the nested sampler for \emph{longer} would not fix this, because
the run stops when the live points hold less than $10^{-3}$ of the evidence, so more
iterations would just add more points with very small weights.
Instead, we would need to increase the number of live points until the modes are well
sampled, which can be expensive for complex posterior pdf structure like we see here.

To test this, we also run the nested sampler five times with 10,000 live points.
The samples then look much more like the rejection samples (right panel of
Figure~\ref{fig:nested-sampling}), and the fraction of the posterior probability between
12.5 and $15.5~\unit{day}$ ranges from 0.122 to 0.135, much closer to the range of the
rejection runs.
However, the uncertainty of each run is now about 0.002, so the runs still differ by
more than their uncertainties.
The fraction of samples below $10~\unit{day}$ also varies as much as with 1000 live
points (0.315 to 0.349 compared to 0.327 to 0.338 for the rejection runs).
We use four representative period ranges (below 10, 12.5--15.5, 22.5--25.5, and above
100 days) to compare the scatter of the fractions around the mean of the rejection runs
with the scatter expected from the effective sample size.
With this estimate, a run with 1000 live points returns about 330 independent samples,
and a run with 10,000 live points returns about 2100 (compared with effective sample
sizes of about 2200 and 23,000, respectively).

The rejection sampler is much faster for this dataset.
On a laptop CPU (Apple M4 Max), after \package{JAX} compilation, one rejection run for
this source takes 1.6~\unit{s} and returns about 9100 independent samples, whereas one
nested sampling run with 1000 live points takes 3.5~\unit{s} and returns weighted
samples with an effective sample size of about 2200.
So, \harv's rejection sampler therefore generates about 5700 effective samples per
second versus about 640 per second for the nested sampling run with 1000 live points.
But the nested sampling number is too optimistic, because the effective sample size of
the weights does not include the scatter of the mode probabilities we find above (see
previous paragraph).
With 10,000 live points, one run takes 33~\unit{s} and produces about 2100 independent
samples, or about 65 per second.
Even if the number of independent samples grew in proportion to the number of live
points beyond 10,000, matching the precision of one rejection run would need about
40,000 live points, and the nested sampler would be about 90 times slower than the
rejection sampler.

For pipelines that analyze many sources with varied data quality or number of epochs, we
therefore recommend using the rejection sampler first, and then using nested sampling or
MCMC only for sources that need follow-up.

\subsection{Case study 4: Population inference with simulated data}
\label{sec:cases-pop-inf}

The first three case studies focus on analyzing individual sources, or small numbers of
sources.
But the main motivation for \harv\ is to enable population inferences that use all
sources from a large sample, whether or not their orbits are well constrained
(Section~\ref{sec:intro}).
In this case study, we demonstrate this use case by inferring the binary fraction and
population distributions of period, eccentricity, and companion mass for a simulated
population.
Here we use simulated data so we can compare the inferred population with truth.

\begin{figure*}
  \centering
  \includegraphics[width=\textwidth]{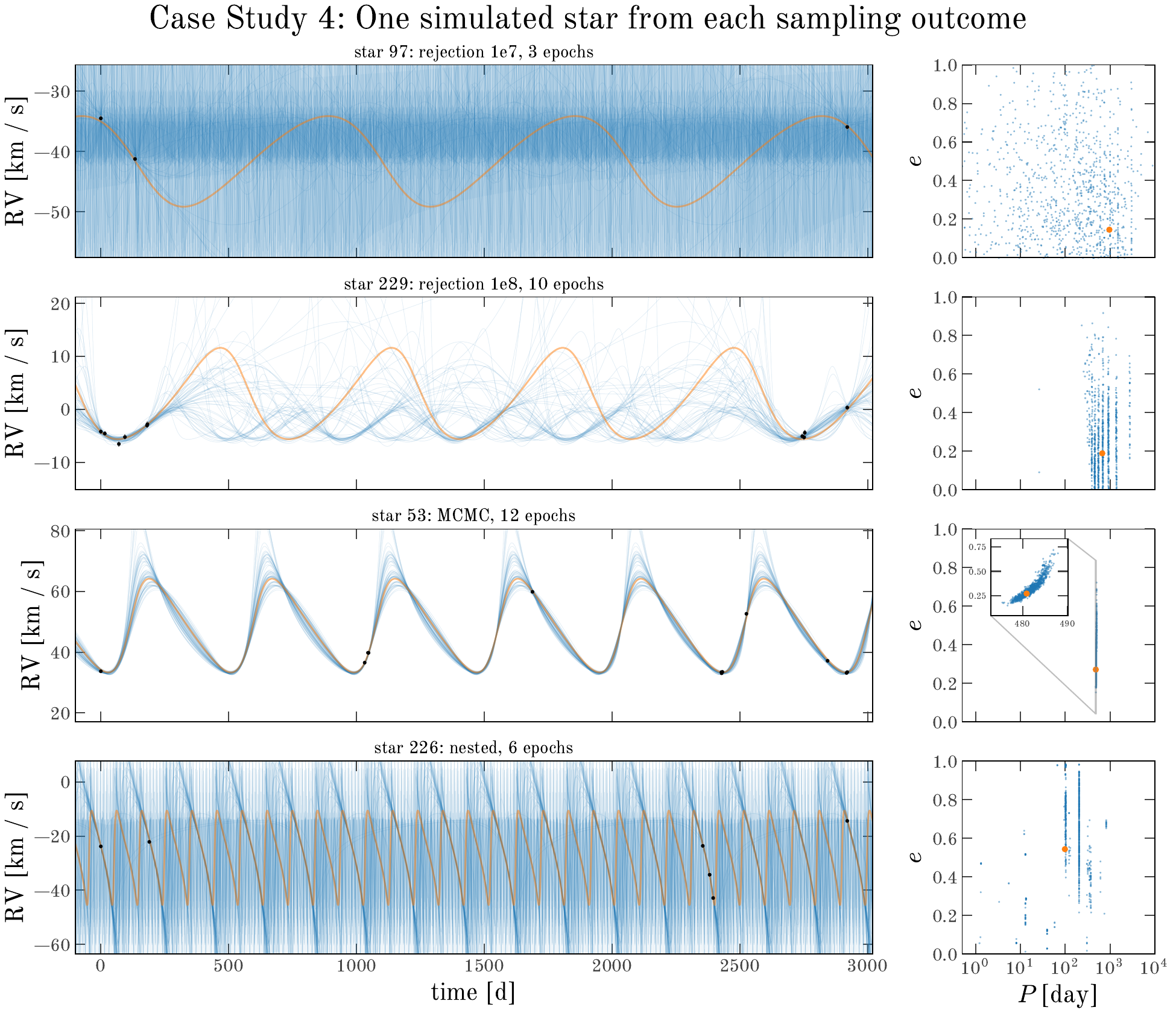}
  \caption{
    Simulated data and posterior samples for one star (each row shows data and results
    for one star) from each outcome of the sampling procedure in
    Section~\ref{sec:cases-pop-inf}.
    \textit{Left:} This panel shows the simulated RV data (black markers), orbits
    computed from 128 posterior samples (blue lines), and the true orbit (orange line).
    \textit{Right:} This panel shows a projection of the posterior samples, showing
    period and eccentricity of all 1024 posterior samples (blue points) over the range
    of the interim prior.
    The larger (orange) markers show the true values.
    The NUTS samples (third row) are in one narrow mode, so we zoom in with an inset
    panel.
    \label{fig:pop-rv-examples}
  }
\end{figure*}

We simulate RV data for 256 stars, 102 ($\approx 40\%$) of which have a companion, each
observed between 3--12 times (drawn uniformly) over 8 years with uncertainties of
$0.5~\unit{km.s^{-1}}$.
For the binaries, we sample the log period $\log_{10}(P/\unit{day})$ from a normal
distribution with mean 5.03 and standard deviation 2.28 \citep{Raghavan:2010}, truncated
to periods between 5 and $1000~\unit{day}$.
We sample the eccentricity $e$ from a normal distribution with mean 0.4 and standard
deviation 0.15, truncated to $[0, 1]$.
Finally, we sample the log companion mass $\log_{10}(m_2/\unit{\Msun})$ from a normal
distribution with mean $\log_{10} 0.9$ and standard deviation 0.3, truncated to
companion masses between 0.1 and $1~\unit{\Msun}$.
We sample primary masses $m_1$ from a normal distribution with a mean $1~\unit{\Msun}$
and standard deviation 0.3, truncated to $[0.5, 1.5]$.
These choices are very roughly meant to represent a population of solar-type stars with
low-mass companions, but we do not attempt to match any particular population correctly
or in detail.
We generate the orientation angles for the orbits to be isotropic.
The resulting velocity semi-amplitudes in the simulated sample vary from $\approx 0.7$
to $74~\unit{km.s^{-1}}$.
For each star, we also simulate a measurement of $m_1$ with an uncertainty of
$0.05~\unit{\Msun}$.
Figure~\ref{fig:pop-rv-examples} (left column, black markers) shows the simulated RV
data for four stars from this sample (error bars are plotted, but they are comparable to
the marker size).

For the inference pipeline, we start by running the \harv\ rejection sampler on each
simulated system.
We use the same prior pdf for all sources, which is the interim prior of the
hierarchical model below.
This prior pdf is uniform in $\ln P$ between 0.5 and $10^4~\unit{day}$, a truncated
normal distribution for $e$ with mean 0.1 and standard deviation 0.4, uniform in the
angles or phases $\omega$ and $\tau$, $\sigma_{K,0} = 30~\unit{km.s^{-1}}$, and
$\sigma_{v_0} = 50~\unit{km.s^{-1}}$.
As in Section~\ref{sec:cases-apogee}, we draw $J = 10^7$ prior samples once and run the
rejection sampler on every star using these samples; we keep at most 1024 samples per
star.
Because every star's samples come from the same library of prior samples, the Monte
Carlo errors in the per-star sums of Equation~\ref{eq:pop-mixture} are correlated
between stars.
This does not bias the sum for any one star, but the variance of the population
log-likelihood can be larger than the sum of the per-star variances.
After this stage, 182 of 256 sources return 1024 samples, 40 sources return $\leq3$
samples, and 34 sources return an intermediate number of samples.
For stars with fewer than 1024 samples, we follow up the rejection sampling with other
strategies.
First, we rerun the rejection sampler with a library of $10^8$ prior samples that we
again generate once and reuse for all such cases.
If, after this, a star gets 1024 samples, we save the samples and proceed.
If instead, after rejection sampling with $10^8$ prior samples, a source still has fewer
than 1024 samples and all samples are within one period mode, we continue with NUTS
(again, we run four chains initialized from the rejection samples) and keep 1024 MCMC
samples if the runs converge with $\hat{R} < 1.2$.
If a source has fewer than 1024 samples after the larger rejection sampling run and
samples are spread across multiple period modes, or if the NUTS run leads to $\hat{R}
\geq 1.2$, we run a nested sampler following Section~\ref{sec:cases-interop} with
\num{10000} live points and downsample the generated samples until we have 1024.
After this procedure, all 256 sources have 1024 posterior samples: 182 from the first
rejection sampling run with $10^7$ prior samples, 11 from the run with $10^8$ prior
samples, 30 from NUTS, and 33 from nested sampling (26 with samples in several period
modes and 7 for which NUTS did not converge).
Figure~\ref{fig:pop-rv-examples} shows the posterior samples for one star from each of
these outcomes.

We model the population as a mixture of binaries and stars with a constant velocity.
The mixture weight for this model is the binary fraction $f_{\rm bin}$.
However, a ``binary fraction'' is not a useful concept until we define what we mean by
a binary.\footnote{Technically, there are no single stars: every star presumably has
planets, asteroids, or at least a few dust grains in orbit around it, so the
(ambiguously-defined) ``binary fraction'' of any population is 1 (which we can assert
without looking at any data).
In practice, a binary fraction counts the stars with a companion that is massive enough
and close enough to matter for some data set.
However, these thresholds are often set implicitly by the data and the method rather
than explicitly stated.
An implication of this is that two binary fractions measured from different surveys of
the same stars can therefore disagree even though both may be technically correct.
So an unqualified ``binary fraction'' is not a measurement of anything in particular.
Here, we simulate the data to be either exactly single stars or stars with a
companion in a specific range of period and mass, so our pedantic objections are not
really relevant for this experiment.}
We define $f_{\rm bin}$ as the fraction of stars that have a companion with properties
in the range period $5 < P < 1000~\unit{day}$ and mass $0.1 < m_2 < 1~\unit{\Msun}$.
All other stars count as stars without a companion (as we have defined companion).
In this toy simulation, every companion is in this range, so we can model the other
stars as having constant velocity.
With real data, stars with a companion outside the range would need another component in
the mixture model.
The likelihood for the data $\vec{y}_n$ for star $n$ is
\begin{multline}
  p(\vec{y}_n \given f_{\rm bin}, \vec{\alpha}) = \\
    f_{\rm bin}\,\frac{Z_{{\rm int},n}}{S_n} \sum_{k=1}^{S_n} r_{nk}(\vec{\alpha})
    + (1 - f_{\rm bin})\,Z_{{\rm single},n}
  \label{eq:pop-mixture}
\end{multline}
where the first term is the reweighting sum of Equation~\ref{eq:reweight} with
$w_{nk} = 1/S_n$, each star
$n$ has $S_n$ samples from the procedure above (here, all $S_n = 1024$ but in general
they could be different), $Z_{{\rm int},n}$ is the evidence for the binary model under
the interim prior (from the rejection sampler), the $r_{nk}(\vec{\alpha})$ are the
prior ratios for each sample (defined below in Equation~\ref{eq:pop-weights}), and
$Z_{{\rm single},n}$ is the evidence for the constant-velocity model, which we compute
as in Section~\ref{sec:cases-apogee}.
Unlike in Equation~\ref{eq:reweight}, the evidences do not cancel here because the two
terms in the mixture are different models.

\begin{figure*}
  \centering
  \includegraphics[width=\textwidth]{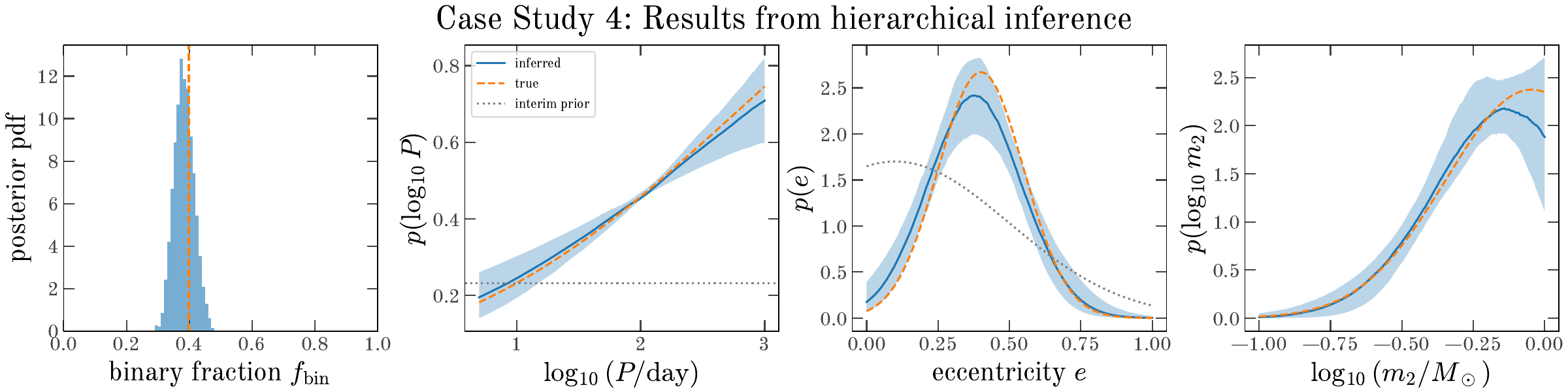}
  \caption{
    Results from a population inference using all 256 simulated stars
    (Section~\ref{sec:cases-pop-inf}).
    \textit{Left:} The posterior pdf of the binary fraction $f_{\rm bin}$, with the true
    value (vertical dashed line).
    \textit{Other panels:} The population distributions of $\log_{10} P$, $e$, and
    $\log_{10} m_2$ for the binaries.
    The blue lines and bands show the median and the 16th to 84th percentiles at each
    point over 512 posterior samples of $\vec{\alpha}$.
    The dashed orange lines show the true distributions.
    The dotted gray lines show the interim prior.
    \label{fig:population}
  }
\end{figure*}

For the binaries, the population model is a truncated normal distribution in each of
$\log_{10} P$ (between 5 and $1000~\unit{day}$), $e$ (between 0 and 1), and $\log_{10}
m_2$ (between 0.1 and $1~\unit{\Msun}$), so $\vec{\alpha}$ contains the three means and
three standard deviations of the normal distributions, and $f_{\rm bin}$ is the fraction
of stars with a companion in this range of $P$ and $m_2$.
Now, the samples are in $(P, e, K)$, but we express the population model in terms of
$(P, e, m_2)$.
So, for each sample, we draw $m_1$ from its measurement and $\cos i$ from a
uniform distribution, solve the binary mass function for $m_2$, and include the Jacobian
of the transformation from $K$ to $m_2$ in the prior ratio
\begin{equation}
  r_{nk}(\vec{\alpha}) =
    \frac{p(P_{nk}, e_{nk}, m_{2,nk} \given \vec{\alpha})}{p(P_{nk}, e_{nk}, K_{nk} \given \vec{\alpha}_0)}
    \left|\pderiv{m_2}{K}\right|_{nk}
    \label{eq:pop-weights}
  \quad ,
\end{equation}
where the denominator is the interim prior pdf, and $r_{nk} = 0$ outside the range of
$P$ and $m_2$.
We sample $f_{\rm bin}$ and $\vec{\alpha}$ with NUTS using uniform priors for all
parameters.

Equation~\ref{eq:pop-mixture} uses the evidence estimates from the rejection sampler,
which are not reliable when $J_{\rm eff} \approx 1$ (Section~\ref{sec:cases-interop}).
However, this does not affect the results here.
The 23 stars with $J_{\rm eff} < 3$ all have $\ln \mathcal{B} = \ln Z_{\rm int} - \ln
Z_{\rm single} > 1400$, so the binary term in Equation~\ref{eq:pop-mixture} vastly
dominates, and $Z_{{\rm int},n}$ is only a constant factor in the likelihood.
The 159 stars for which the two models are comparable ($|\ln \mathcal{B}| < 5$) all have
$J_{\rm eff} > 10^5$, and for these the evidence estimates are precise.

Figure~\ref{fig:population} shows the results of the population inference.
We find that the inferred binary fraction is $f_{\rm bin} = 0.39 \pm 0.03$, compared
with the true value 0.40.
We infer that the eccentricity distribution has mean $0.38 \pm 0.03$ and standard
deviation $0.17 \pm 0.03$ (true values 0.4 and 0.15), and the distribution of $\log_{10}
m_2$ has mean $-0.13^{+0.23}_{-0.10}$ and standard deviation $0.26^{+0.13}_{-0.08}$
(true values $-0.05$ and 0.3).
The mean and standard deviation of the period distribution are degenerate because the
true mean is outside the range of $P$, so the truncated distribution is close to an
exponential in $\log_{10} P$.
However, the inferred distribution over the range where we have data agrees with the
true distribution.
All 256 stars enter this inference, including the 154 stars that do not have a companion
and the binaries with poorly constrained orbits.
In other words, we do not have to select stars by the signal-to-noise ratio of the
inferred orbital parameters or by the detection significance.

\section{Performance benchmarks}
\label{sec:performance}

Here we compare the speed of the rejection sampler in \harv\ with \joker, and we also
report some computational cost scaling curves for the rejection sampler as a function of
the signal-to-noise of the data.
We run all of these benchmarks in the same environment on a Linux workstation using
either one core of an Intel Xeon w5-3435X CPU or one NVIDIA RTX 6000 Ada GPU.
The likelihood evaluations for different prior samples are independent, so with several
devices the times should divide by the number of devices, modulo overheads for
distributing tasks and collecting or saving the results.

\subsection{Comparison to \joker}

We time the rejection sampler implemented in \harv\ versus \joker\ (version 1.3.1) on
simulated RV data, using the prior from Section~\ref{sec:cases-apogee} in both codes.
The simulated RV orbit has $P = 200~\unit{day}$, $e = 0.3$, and $K =
10~\unit{km.s^{-1}}$, observed with uncertainties $0.5~\unit{km.s^{-1}}$ over 3 years
(and a variable number of observations).
The cost per prior sample does not depend on the values of the data, so the choices of
the simulated orbit and the observation times do not affect these benchmarks (only the
number of observations matters for scalings).
While both \harv\ and \joker\ can stream cached prior samples from disk, here we only
benchmark a situation where the full prior samples cache is already in memory.
For \harv, we run each benchmark three times and report the median time, so we
neglect \jax\ compilation overheads.
For \joker\ likelihood evaluations we time a single call (there is no compilation
overhead).
We also do not include the time to generate the prior samples for each code, which we
only do once at the start (but even this is much faster in \harv\ than in \joker).

As a first test, before we run the benchmarks, we check that the two rejection samplers
are consistent.
We use RV data from the source 2M02323271+3350542 (Section~\ref{sec:cases-apogee}),
which has three observations, and we run both rejection samplers with $J = 10^6$ prior
samples.
The two codes use different conventions for the orbital phase (\joker\ uses the mean
anomaly at a reference time and \harv\ the phase of pericenter), so instead of comparing
the marginal likelihoods sample by sample, we compare the evidence estimates from
independent sets of prior samples drawn by each code.
We find for \harv\ $\ln p(\vec{y}) = -8.490 \pm 0.002$ and for \joker\ $-8.487 \pm
0.002$, which are consistent within the Monte Carlo uncertainty.

\begin{figure*}
  \centering
  \includegraphics[width=\textwidth]{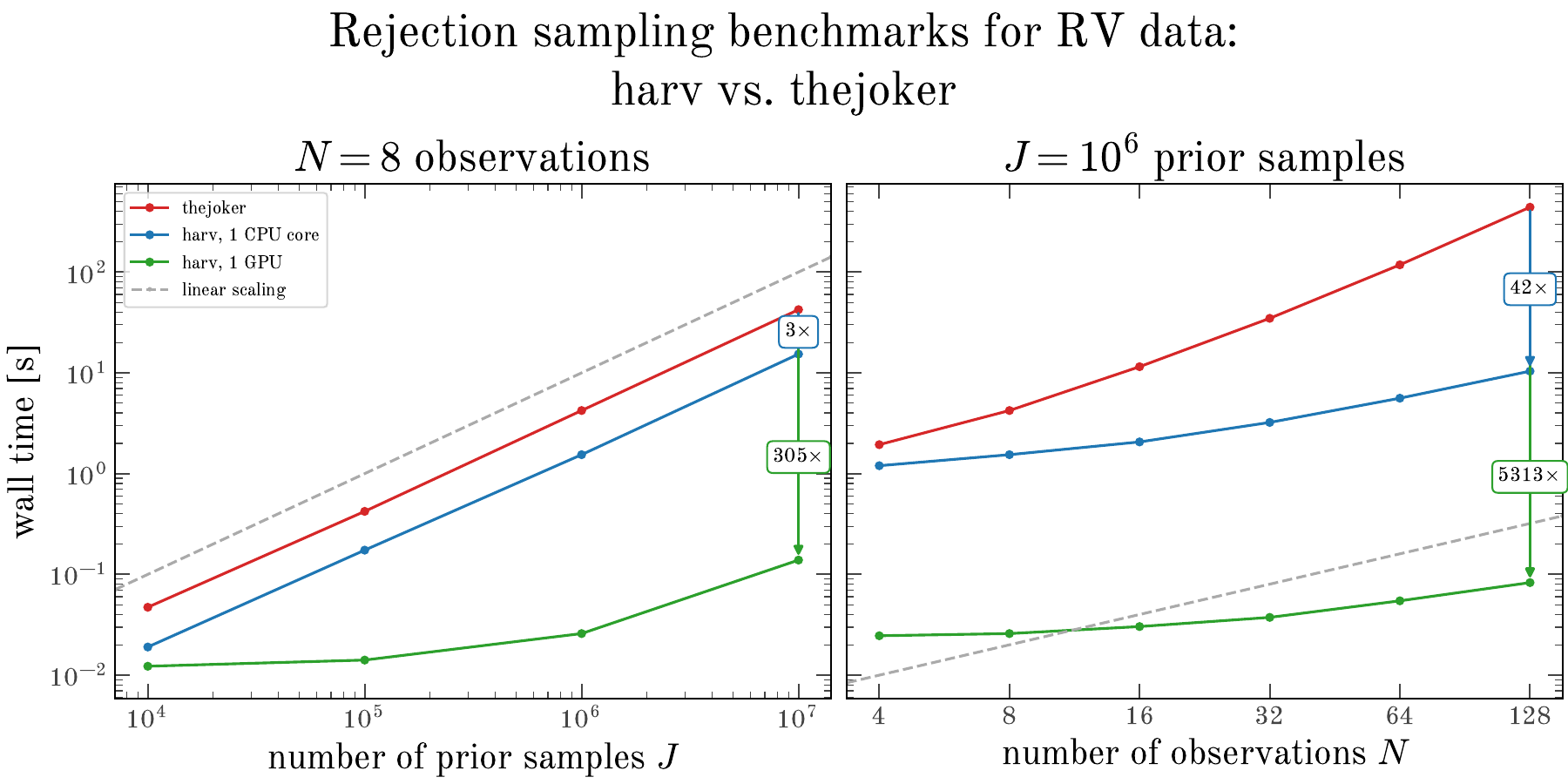}
  \caption{
    Wall time of the rejection sampler as a function of the number of prior samples $J$
    (left, for $N = 8$ observations) and as a function of the number of observations $N$
    (right, for $J = 10^6$).
    Two lines (red and blue) show runs of \joker\ and \harv\ on only one CPU core,
    respectively, and the third line (green) shows \harv\ on one GPU.
    In the right panel, the wall time in seconds equals the time per prior sample in
    microseconds (we use $10^6$ prior samples).
    The vertical arrows mark the speed-up of \harv\ over \joker\ at $J = 10^7$ (left)
    and $N = 128$ (right).
    \label{fig:bench-thejoker}
  }
\end{figure*}

Next, we run the benchmarks for different numbers of prior samples $J$ and different
numbers of observations $N$.
Figure~\ref{fig:bench-thejoker} shows the total wall time for each code to run the
rejection sampler given $J$ prior samples (left panel) and $N$ observations (right
panel).
On the CPU, with $N = 8$, the time grows linearly with $J$ for both packages, so the
time per prior sample is nearly constant.
On the GPU, there is a fixed overhead of about 13~\unit{ms} per call that
dominates the runtime below $J \approx 10^5$, but the wall time approaches linear in $J$
for larger $J$.
At $N = 8$ and $J = 10^7$, \joker\ takes 4.2~\unit{\micro\second} per prior
sample, and \harv\ takes 1.5~\unit{\micro\second} on one CPU core and
0.014~\unit{\micro\second} on one GPU, so \harv\ is 2.8 and 310
times faster.

For small numbers of observations per source, \harv\ is only a few times faster than
\joker\ on one CPU core, but the speed-up grows substantially for data with more
observations.
For each prior sample, \joker\ builds the $N \times N$ matrix $\mat{C} +
\mat{X}\,\mat{\Lambda}\,\transp{\mat{X}}$ of Equation~\ref{eq:marginal-likelihood} and
its inverse, and computes an LU factorization for the determinant, so the cost per
sample grows faster than linearly with $N$ (roughly as $N^2$, we see empirically).
\harv\ does not explicitly create these matrices.
When the data covariance $\mat{C}$ is diagonal, we instead use the matrix inversion and
determinant lemmas (Section~\ref{sec:methods-marginal}), which only need $Q \times Q$
matrices and cost $\mathcal{O}(N\,Q^2)$ per prior sample.
On one CPU core, the time per prior sample therefore scales linearly with $N$, so
\harv's speed-up relative to \joker\ grows with $N$.
As we can see in Figure~\ref{fig:bench-thejoker}, at $J = 10^6$ prior samples, the
factor speed-up is $\approx 1.6$ for $N = 4$ but increases to $\approx 42$ for $N = 128$
on one CPU core.
Another advantage of \harv\ is that it can run on a GPU, which is much faster than
either code on a CPU.
However, for large samples of stars on a cluster with many CPU cores, it may still be
faster to run \harv\ on the CPU cores.
This is the case when the number of available CPU cores divided by the number of
available GPUs is larger than the speed-up factor of \harv\ on one GPU versus one CPU
core, at the typical number of observations per star in the sample.

\subsection{Rejection sampler benchmarks for varied signal-to-noise}

\begin{figure*}
  \centering
  \includegraphics[width=\textwidth]{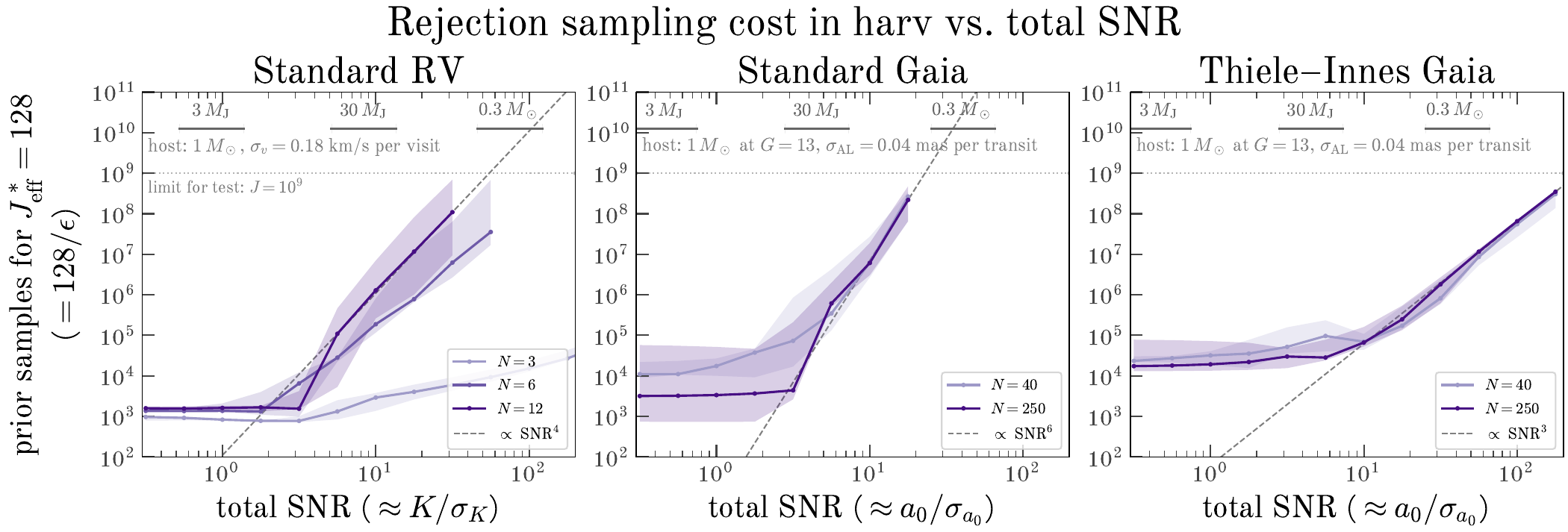}
  \caption{
    The number of prior samples needed to get $\targESS = 128$ effective samples from
    the posterior as a function of the total SNR of the orbit.
    We fix the uncertainty of each observation at the motivated values given in the
    text, and the amplitude of the orbit then sets the total SNR.
    We arbitrarily pick a $1~\unit{\Msun}$ host star for the \gaia\ astrometry and an
    apparent magnitude $G = 13$.
    The (purple) lines show the median and range from three realizations of the
    simulated data, and we show a few choices for the number of observations $N$.
    The dashed line is proportional to (total SNR)$^D$ where $D$ is the number of
    sampled parameters for each parameterization.
    The horizontal gray line at $10^9$ is the largest number of prior samples we
    consider (the curves end where even $10^9$ prior samples give $J_{\rm eff} < 128$).
    Bars at the top show ranges of the total SNR for the three representative companions
    for different numbers of observations.
    \textit{Left}: The standard RV parameterization, with $D = 4$ sampled parameters.
    \textit{Middle}: The standard \gaia\ parameterization, with $D = 6$ sampled
    parameters.
    \textit{Right}: The Thiele--Innes \gaia\ parameterization, with $D = 3$ sampled
    parameters.
    In both \gaia\ panels, we marginalize the parallax.
    \label{fig:bench-ess}
  }
\end{figure*}

Timing the rejection sampler for different prior sample cache sizes and number of
observations is useful for understanding the performance of \harv\ in different regimes,
but in real applications we usually care about the time it takes to return a given
number of effective samples from the posterior.
For example, imagine we want to require that we get some number $\targESS$ of effective
samples from the posterior for each source in a large sample of stars.
This time depends strongly on the data, because more informative data lead to a
smaller posterior volume and therefore a smaller efficiency $\epsilon = J_{\rm eff}/J$
of the rejection sampler.
To do this with the rejection sampler (which we may not want to enforce, but let's
imagine we do for now), we need to draw $J$ prior samples, evaluate the likelihood for
each, and then compute the effective sample size $J_{\rm eff}$ for the samples.
Once $J$ is large enough to resolve the posterior pdf, $J_{\rm eff}$ grows in proportion
to $J$, so in this regime the efficiency $\epsilon$ depends on the data and the prior
pdf, but not on $J$.
Getting $\targESS$ effective samples therefore takes $\targESS/\epsilon$ prior samples,
or a time
\begin{equation}
  t(\targESS) \approx \frac{\targESS}{\epsilon \, R}
  \quad ,
\end{equation}
where $R$ is the number of prior samples evaluated per second.
$R(N)$ depends on the number of observations, the parameterization, and the device, but
not on the values of the data.
The efficiency $\epsilon$ depends on the data: $\epsilon \lesssim e^{-\mathcal{H}}$,
where $\mathcal{H}$ is the information over the sampled parameters
(Section~\ref{sec:methods-scaling}).

For informative data ($\mathcal{H}$ above some threshold), the posterior width in each
sampled parameter scales with $1/{\rm SNR}$, where SNR is the total signal-to-noise
ratio of the orbit of the observed star,
\begin{equation}
  {\rm SNR}^2 = \sum_{n=1}^{N} \left(\frac{s_n}{\sigma_n}\right)^2
  \quad ,
\end{equation}
where $s_n$ is the orbital signal at observation $n$ (after subtracting the model
without a companion --- i.e., constant velocity for RV data or the five parameter
astrometric solution for astrometry), and this is the total SNR over all $N$
observations, not the signal-to-noise ratio of a single visit or transit.
From the Fisher information, when the number of observations or the total SNR is large,
the posterior volume within a mode over the $D$ sampled parameters scales like ${\rm
SNR}^{-D}$, so we expect $\epsilon \propto {\rm SNR}^{-D}$.
The number of sampled parameters is $D = (4, 6, 3)$ for the three parameterizations we
consider here: the standard RV parameterization, the standard \gaia\ parameterization,
and the Thiele--Innes \gaia\ parameterization, respectively (the nonlinear parameters in
Table~\ref{tab:parameterizations}), when the parallax is marginalized.

We test these scalings numerically with simulated data.
We use the prior from Section~\ref{sec:cases-apogee} for RVs and the prior from
Section~\ref{sec:cases-gaia4} (without jitter) for \gaia\ astrometry.
For both \gaia\ parameterizations, we compute the prior widths on $a_0$ (or the
Thiele--Innes constants) and on the proper motion from a fixed parallax, so that the
parallax is marginalized (Section~\ref{sec:methods-priors-nonlinear}).
For the Thiele--Innes parameterization, we use the same priors, with a Gaussian prior on
each Thiele--Innes constant that has the same width as the prior on $a_0$.
We do not apply a Jacobian correction, so the implied prior on $a_0$ and the orientation
of the orbit differs from the standard parameterization.
We use the same simulated data for both \gaia\ parameterizations.
For each data type, we use the same orbit shape but vary its amplitude ($K$ or $a_0$) to
set the total SNR.
For RVs, we use $P = 200~\unit{day}$ and $e = 0.3$ over a baseline of 3 years, and for
\gaia, we use $P = 150~\unit{day}$ and $e = 0.2$ over a baseline of 5 years (with random
scan angles and parallax factors).
For the \gaia\ astrometry, we assume a Sun-like host star with $M=1~\unit{\Msun}$ and $G
= 13$ (corresponding to $\varpi = 2.16~\unit{mas}$ from the distance modulus with $M_{G}
= 4.67$ and no extinction).
The parallax matters for the \gaia\ data because it sets the angular size of the orbit
and the widths of the prior pdfs on $a_0$ and the proper motion.
We pick the periods above to be near the logarithmic middle of the period prior and not
near 1 year, harmonics of 1 year, or the 63-day precession period of the \gaia\ scanning
law.

We fix the uncertainty of each observation at values motivated by \apogee\ and \gaia,
$\sigma_v = 0.18~\unit{km.s^{-1}}$ per visit for \apogee-like RVs and $\sigma_{\rm AL} =
0.04~\unit{mas}$ per transit for \gaia\ at $G = 13$ (the median per-transit uncertainty
at this magnitude in the noise model of \citealt{Yahalomi:2026}), and we set the total
SNR between 0.3 and 1000 by scaling the amplitude ($K$ or $a_0$), using $N = 3$, 6, and
12 RVs and $N = 40$ and 250 \gaia\ transits.
We do this for three realizations of the observation times and noise for each.
For each dataset, we start with $J = 10^6$ prior samples and multiply $J$ by 10 until
$J_{\rm eff} \geq 128$ or $J = 10^9$, and we estimate $\epsilon$ from the last run.
With the uncertainties fixed, we can also mark ranges of SNR that correspond to three
representative companions on these orbits around a $1~\unit{\Msun}$ star: an M dwarf
($0.3~\unit{\Msun}$), a brown dwarf ($30~\unit{\Mjup}$), and a giant planet
($3~\unit{\Mjup}$).

Figure~\ref{fig:bench-ess} shows the results from these benchmarks.
At low SNR, all of the curves are flat but well below $\epsilon = 1$.
Even when the orbit is not detected, the data are precise enough to rule out large
orbits for most periods and orientations, which is informative compared with the broad
prior pdf on the amplitude.
The flat level depends on the parameterization and data.
Above a total SNR of about 3 (RVs and standard \gaia) or 10 (Thiele--Innes), the cost
(i.e. the number of prior samples we need to reach a certain number of effective
samples) rises steeply.
In each panel, we plot a dashed line proportional to (total SNR)$^D$, where $D$ is the
number of sampled parameters.
The measured cost scales roughly as expected with the total SNR, except for the standard
RV parameterization with $N=3$ or 6 observations (left panel).
With so few observations, the model is under-constrained, so the cost grows more slowly
with the total SNR.

\section{Discussion}
\label{sec:discussion}

We designed \harv\ to be a general-purpose tool for modeling the Keplerian orbits of
companions around stars, and for using the resulting posterior samples to perform
population inferences.
However, the current implementation has some limitations that we discuss here (some of
which result from our assumptions in Section~\ref{sec:design-assumptions}).

\subsection{General limitations and missing functionality}
\label{sec:discussion-general}

\harv\ currently supports only one Keplerian companion per model, so we cannot generally
model multi-planet or multiple-star systems, which are fairly common.
We can model these cases if there is a clear separation of timescales between the
orbits, so that the inner orbit is effectively an unperturbed Keplerian orbit and the
outer orbit appears as a polynomial trend in the data.

In several places, \harv\ also assumes that the companion is dark.
An exception in the current implementation is double-lined (SB2) RV systems, in which
both components are luminous and have their own RVs (this is supported).
However, \harv\ does not model the flux ratio of a luminous companion, which sets the
photocenter motion in the astrometry, and in a joint model of RVs and astrometry it
assumes that the photocenter traces the orbit of the star with the measured RVs
(Equation~\ref{eq:conv-tie}), which is not true when the companion is luminous.

\harv\ does not yet include ``perspective acceleration'' in the astrometric model, which
matters for very nearby or fast-moving stars.
This effect arises because uniform motion in 3-space does not project to uniform motion
on the sky when the distance to the star or its direction changes significantly over the
time baseline of the observations.

Finally, \harv\ supports only one-dimensional \gaia\ epoch astrometry, and not (yet)
relative or absolute astrometry from other instruments.
We would like to add all of these features in future versions of \harv --- contributions
are welcome.

\subsection{Limitations of the rejection sampler}
\label{sec:discussion-rejection}


The main limitation of the rejection sampler approach is computational cost for sources
with informative data.
The rejection sampler needs a sampling of the prior pdf that is dense compared to the
structure of the posterior pdf, and the number of prior samples this takes grows
exponentially with the information $\mathcal{H}$ (Section~\ref{sec:methods-scaling}).
For informative data, $\mathcal{H}$ grows with the logarithm of the total
signal-to-noise ratio, so the cost grows as a power of the total signal-to-noise ratio
with an exponent set by the number of sampled parameters
(Section~\ref{sec:performance}).
For sparse or noisy data, which is where the rejection sampler is most useful, this is
cheap.
In Section~\ref{sec:cases-pop-inf}, 182 of the 256 simulated stars return 1024
samples from $10^7$ prior samples.
For informative data, it quickly becomes prohibitive: 2M20183197+1953430
(Section~\ref{sec:cases-apogee}) has $\mathcal{H} \approx 26$, so resolving its
posterior pdf would take $\sim 10^{11}$ prior samples (with $10^8$ only one sample
survives and is not guaranteed to be near the dominant mode).
For a large survey, we therefore expect most of the computing time to go to the minority
of sources that need follow-up with other methods.
In Section~\ref{sec:cases-pop-inf}, the 33 stars that needed nested sampling (13\% of
the sample) took about half of the total computing time, whereas the 182 stars that
needed only the first rejection run took only about 20\% of the total time.

A second limitation is that when only one or a few samples survive, they may not be in
the dominant mode of the posterior pdf, and the diagnostics computed from them cannot
reveal this (Section~\ref{sec:methods-gmax}).
A single sample is, of course, ``unimodal'' in period, and MCMC chains that all start
from a single sample will converge to the same mode with $\hat{R} \approx 1$, as they
did for 2M20183197+1953430 (Section~\ref{sec:cases-apogee}).
Any automated pipeline that makes follow-up decisions based on the number of samples
returned by the rejection sampler should therefore also check the goodness of fit, such
as the $\chi^2$ of the best sample.
But this check assumes that the model describes the data (which it may not, because of
the limitations in Section~\ref{sec:discussion-general}) and that the uncertainties are
well calibrated.
These issues are not specific to the rejection sampler, and any pipeline that instead
uses optimization or MCMC directly for sources with many epochs or a high total
signal-to-noise ratio will have the same issues.

The hardest sources are ones where the rejection sampler returns only a few samples, and
those samples are spread over several period modes (2M15035796-4533457 in
Section~\ref{sec:cases-apogee}, and 26 of the simulated stars in
Section~\ref{sec:cases-pop-inf}).
Within the rejection sampler, the only path forward is to generate and run with more
prior samples.
The other option is nested sampling or another method that can handle multimodality,
but this is not a simple drop-in replacement, and its settings need tuning to make sure
that the posterior pdf is well sampled.
We saw this in Section~\ref{sec:cases-interop}, where nested sampling with too few live
points fails to resolve the posterior pdf.

\subsection{Limitations of population modeling with per-source samples under an interim prior}
\label{sec:discussion-population}

The hierarchical model (Equation~\ref{eq:pop-marginal}) has a per-source
marginalization integral that we replace with a Monte Carlo sum over the posterior
samples from the rejection sampler.
This is what lets us use every source, but it also means that the population inference
can depend on the quality of the per-source samples in ways that are not obvious.

First, the interim prior has to be nonzero (i.e., have support) everywhere the
population model is nonzero, and in practice it also has to put enough probability mass
in those regions.
By ``enough,'' we mean that the effective sample size of the terms in the Monte Carlo
sum (Equation~\ref{eq:reweight}) must be large enough to make the sum precise.
If this is not the case, the population inference can be biased without any warning.
This is why we choose an interim prior in Section~\ref{sec:cases-pop-inf} that covers a
broader period range (0.5 to $10^4~\unit{day}$) than the population model (5 to
$1000~\unit{day}$).
However, a broad interim prior has a cost, in that many of the posterior samples will
have effectively zero weight in the population model (and are therefore wasted).

Second, each sum is a Monte Carlo estimate and uses a finite number of samples, so the
population likelihood is itself noisy, and the amount of noise depends on the number of
samples and the effective sample size of the weights in each sum.
The effective sample size is large when the population model is broader than the
posterior pdf of an individual source.
The exact number of samples needed depends on the population model, the interim prior,
and the posterior pdf of each source (and variations of, withing a sample of sources).
However, some general guidance has been developed in the context of gravitational-wave
population inference with a similar setup \citep{Essick:2022}.
It turns out that what matters most is the noisiness of the log-likelihood difference
between values of $\vec{\alpha}$ where the posterior pdf of $\vec{\alpha}$ is large.
Meaning, if the log-likelihood is noisy at some arbitrary value of $\vec{\alpha}$ but
smooth in the region where the posterior pdf of $\vec{\alpha}$ is largest, then this
variance does not matter.
Because the posterior pdf of $\vec{\alpha}$ becomes narrower as the number of sources
grows, the number of samples we need per source does not grow with the number of
sources, unless the population model has features that are narrow compared to the
posterior pdfs of individual sources that are most constraining \citep{Essick:2022}.

In the population inference of Section~\ref{sec:cases-pop-inf}, the effective sample
size of the weights evaluated at the true population parameters is below 10 for 100 of
the 253 stars that have samples in the range of $P$ and $m_2$ (99 of these are stars
without a companion).
This is potentially concerning, but there are some ways to check whether this matters.
The most direct test is to rerun the population inference with independent sets of
per-source samples, or with more samples per source, and confirm that the posterior pdf
over $\vec{\alpha}$ is consistent.
In the population inference case study above, we decided to keep 1024 samples per
source, but we could have kept or required more (e.g., all prior samples that survived
the rejection sampling).
\citet{Essick:2022} recommend a simpler test, which is to draw pairs of values of
$\vec{\alpha}$ from the inferred posterior pdf, estimate the Monte Carlo variance of the
difference in the log-likelihood from the sample variances and covariances of the
weights from each source, and confirm that it is much smaller than 1.
This variance would also tell us which sources need more samples.
Here, those would most likely be the stars without a companion, which are also the ones
that are cheapest to sample, so we should have kept many more than 1024 samples for
those stars (and it would not impact the computational cost of the full population
inference).

Third, problems with the per-source samples from the rejection sampler or the
follow-up sampling can bias the population inference.
For example, a source with samples that are stuck in the wrong mode
(Section~\ref{sec:discussion-rejection}) may enter the population model as a very
precisely constrained orbit at the wrong parameter location.
If a small fraction of stars fail in this way, this will not matter much in practice.
However, these failures are typically not random, as they are more likely for
informative data with periods that are short compared to the baseline, where the period
modes are especially narrow.
So in a large survey they could bias the inferred population systematically.

Finally, as we motivated in the introduction (Section~\ref{sec:intro}), the main
advantage of this framework for hierarchical modeling is that it does not require
modeling the detection efficiency of an orbit-fitting pipeline.
But, of course, it is still dependent on the selection function of the parent sample.
For example, an unresolved binary is brighter than a single star of the same primary
mass, so a magnitude-limited sample will over-represent binaries.
There are many other subtleties in modeling systems with a range of distances, masses,
and luminosities (e.g., when systems are resolved in some cases or unresolved in others,
or when the companion is luminous enough to affect the astrometry or photometry), but
these are not specific to \harv.

\section{Summary}
\label{sec:summary}

In this Article, we introduce and describe \harv, a successor to \joker\
\citep{Price-Whelan:2017} built on \jax\ for inferring the orbits of stellar companions
from radial velocities and \gaia\ epoch astrometry.
The main goal of \harv\ is to enable population inference on large samples of stars with
heterogeneous data quality, including sources with sparse or low signal-to-noise data.
Our main conclusions are:
\begin{itemize}
  \item \harv\ models of time series data are linear in a subset of the parameters
    (e.g., the velocity semi-amplitude and systemic velocity for RV data, the five
    astrometric parameters and the angular semi-major axis for astrometry, or the
    Thiele--Innes constants).
    \harv\ (optionally, but most often) marginalizes over these parameters analytically
    by adopting conjugate normal (or truncated normal) prior pdfs so we only have to
    explicitly sample the nonlinear parameters (Sections~\ref{sec:design} and
    \ref{sec:methods}).
  \item The rejection sampler returns independent posterior samples even if the
    posterior pdf is highly multimodal, and it can be used to estimate the evidence for
    a model.
    Because the models are implemented with \jax, these models also work with
    gradient-based MCMC or fast nested sampling methods for cases where the data are too
    informative to warrant rejection sampling (Sections~\ref{sec:methods} and
    \ref{sec:cases-interop}).
  \item We demonstrate the rejection sampler in four case studies using (1) sparse
    \apogee\ RVs, (2) a joint model of RVs and \gaia\ \dr{4} epoch astrometry for
    Gaia-4b, (3) a comparison of rejection and nested sampling, and (4) a population
    inference on simulated data that uses every star --- including stars without
    companions and stars with poorly constrained orbits --- without enforcing a
    detection threshold (Section~\ref{sec:case-studies}).
  \item The cost of rejection sampling grows exponentially with the information in the
    data, $\mathcal{H}$ (Equation~\ref{eq:information}), whereas the cost of nested
    sampling only grows linearly with the information.
    For informative data, the cost of the rejection sampler scales as a power of the
    total signal-to-noise ratio of the orbit, with the exponent equal to the number of
    sampled parameters (Sections~\ref{sec:methods-scaling}, \ref{sec:cases-interop}, and
    \ref{sec:performance}).
    The rejection sampler is therefore most useful for sparse, noisy, or
    low-signal-to-noise data where other samplers struggle.
    But it is still useful as a first step for inference with more informative data.
  \item On one CPU core, \harv\ is between about 3 and 40 times faster than \joker\
    (for 4 to 128 observations), and the speed-up factor grows with the number of
    observations.
    Its performance scales linearly with the number of prior samples or observations on
    both CPU and GPU, and it can be parallelized across multiple CPU cores or GPUs to
    make survey-wide applications feasible.
    On one GPU, it is about 300 times faster than \joker\ for 8 observations
    (Section~\ref{sec:performance}).
\end{itemize}

\harv\ is open-source software available at \url{https://github.com/adrn/harv} and via
the Python Package Index (\texttt{pip install harv}).
This article uses version \harvversion, and the documentation is at
\url{https://harvey.readthedocs.io/}.
The notebooks and scripts that reproduce the case studies, figures, and tables in this
article are available at \url{https://github.com/adrn/harv-paper}.

\begin{acknowledgments}

  It is a pleasure to thank
  Andy Casey,
  Dan Foreman-Mackey,
  David W. Hogg,
  Victor Ramirez Delgado,
  Sheila Sagear,
  William Thompson,
  and Daniel Yahalomi
  for helpful discussions and feedback during the design and testing of \harv.

  Thanks also to the Scientific Computing Core at the Flatiron Institute for their
  support and assistance with computing resources.
  The Flatiron Institute is a division of the Simons Foundation.

  This work made use of Astropy,\footnote{https://www.astropy.org} a community-developed
  core Python package and an ecosystem of tools and resources for astronomy
  \citep{astropy:2013, astropy:2018, astropy:2022}.

  \paragraph{LLM disclosure}
  I used Claude Opus 4.8, 5.0, and 5.5 during the software development process to
  brainstorm code design and implement features.
  I used Claude Opus 5.5 to brainstorm case study ideas (see
  Section~\ref{sec:case-studies}) to demonstrate the capabilities of \harv\ in this
  article.
  I used Claude Opus 5.5 to review and suggest edits to math and text that I wrote, but
  not to generate any of the math or text from prompts.

  \paragraph{Data acknowledgements}
  Funding for the Sloan Digital Sky Survey IV has been provided by the Alfred P. Sloan
  Foundation, the U.S. Department of Energy Office of Science, and the Participating
  Institutions.

  SDSS-IV acknowledges support and resources from the Center for High Performance
  Computing  at the University of Utah. The SDSS website is www.sdss4.org.

  SDSS-IV is managed by the Astrophysical Research Consortium for the Participating
  Institutions of the SDSS Collaboration including the Brazilian Participation Group,
  the Carnegie Institution for Science, Carnegie Mellon University, Center for
  Astrophysics | Harvard \& Smithsonian, the Chilean Participation Group, the French
  Participation Group, Instituto de Astrof\'isica de Canarias, The Johns Hopkins
  University, Kavli Institute for the Physics and Mathematics of the Universe (IPMU) /
  University of Tokyo, the Korean Participation Group, Lawrence Berkeley National
  Laboratory, Leibniz Institut f\"ur Astrophysik Potsdam (AIP),  Max-Planck-Institut
  f\"ur Astronomie (MPIA Heidelberg), Max-Planck-Institut f\"ur Astrophysik (MPA
  Garching), Max-Planck-Institut f\"ur Extraterrestrische Physik (MPE), National
  Astronomical Observatories of China, New Mexico State University, New York University,
  University of Notre Dame, Observat\'ario Nacional / MCTI, The Ohio State University,
  Pennsylvania State University, Shanghai Astronomical Observatory, United Kingdom
  Participation Group, Universidad Nacional Aut\'onoma de M\'exico, University of
  Arizona, University of Colorado Boulder, University of Oxford, University of
  Portsmouth, University of Utah, University of Virginia, University of Washington,
  University of Wisconsin, Vanderbilt University, and Yale University.

  This work has made use of data from the European Space Agency (ESA) mission {\it Gaia}
  (\url{https://www.cosmos.esa.int/gaia}), processed by the {\it Gaia} Data Processing
  and Analysis Consortium (DPAC,
  \url{https://www.cosmos.esa.int/web/gaia/dpac/consortium}). Funding for the DPAC has
  been provided by national institutions, in particular the institutions participating
  in the {\it Gaia} Multilateral Agreement.

\end{acknowledgments}

\software{
  \texttt{harv} \citep{harv:zenodo},
  \texttt{astropy} \citep{astropy:2013, astropy:2018, astropy:2022},
  \texttt{blackjax} \citep{Cabezas:2024},
  \texttt{equinox} \citep{Kidger:2021},
  \texttt{JAX} \citep{Bradbury:2018},
  \texttt{Jupyter} \citep{Granger:2021},
  \texttt{matplotlib} \citep{Hunter:2007},
  \texttt{numpy} \citep{Harris:2020},
  \texttt{numpyro} \citep{phan2019composable},
  \texttt{pymc} \citep{pymc2023},
  \texttt{thejoker} \citep{Price-Whelan:2017},
  \texttt{tinygp} \citep{Foreman-Mackey:2026},
  \texttt{unxt} \citep{Starkman:2025}.
}

\appendix

\section{Notation and conventions}
\label{apx:conventions}

Table~\ref{tab:notation} lists the symbols that we use in more than one section of the
main text, along with where each is defined or first used.
Symbols that appear in only one section, including those in these appendices, are
defined where they are used.

\begin{table*}
  \centering
  \caption{Symbols used in more than one section of the main text.}
  \label{tab:notation}
  \begin{tabular}{lp{0.5\textwidth}l}
    \hline
    Symbol & Description & Defined / first used \\
    \hline
    \multicolumn{3}{l}{\textbf{Orbit and data}} \\
    $P$, $e$, $\tau$, $\omega$ & period, eccentricity, phase (time of pericenter in units of
      $P$), and argument of pericenter & Eq.~\ref{eq:rv-model} \\
    $K$, $v_0$ & RV semi-amplitude and systemic velocity & Eq.~\ref{eq:rv-model} \\
    $\Omega$, $i$ & longitude of the ascending node and inclination & Eq.~\ref{eq:al-model} \\
    $a_0$ & angular semi-major axis of the photocenter orbit & Eq.~\ref{eq:al-model} \\
    $A$, $B$, $F$, $G$ & Thiele--Innes constants & Eq.~\ref{eq:conv-thiele-innes} \\
    $\varpi$ & parallax & Eq.~\ref{eq:al-model} \\
    $m_2$ & companion mass & Sec.~\ref{sec:methods-priors-nonlinear} \\
    $N$ & number of observations of a source & Eq.~\ref{eq:linear-model} \\
    $T$ & time spanned by the observations of a source & Sec.~\ref{sec:methods-evidence} \\
    \hline
    \multicolumn{3}{l}{\textbf{Linear parameters and marginalization}} \\
    $\vec{y}$ & data for one source ($\vec{y}_n$ for source $n$) & Eq.~\ref{eq:linear-model} \\
    $\vec{\theta}$, $\vec{\beta}$ & nonlinear and linear parameters & Eq.~\ref{eq:linear-model} \\
    $\mat{X}$, $Q$ & $N \times Q$ design matrix, and number of linear parameters
      & Eq.~\ref{eq:linear-model} \\
    $\mat{C}$ & covariance matrix of the data & Sec.~\ref{sec:methods-marginal} \\
    $\vec{\mu}$, $\mat{\Lambda}$ & mean and covariance of the prior pdf over $\vec{\beta}$
      & Sec.~\ref{sec:methods-marginal} \\
    $\sigma_{K,0}$, $\sigma_{a,0}$, $\sigma_{v,0}$, $P_0$ & scales of the prior pdfs over
      $K$, $a_0$, and the proper motion, and reference period
      & Eqs.~\ref{eq:sigma-K}, \ref{eq:sigma-a} \\
    \hline
    \multicolumn{3}{l}{\textbf{Sampling}} \\
    $J$ & number of prior samples & Sec.~\ref{sec:methods-rejection} \\
    $S$, $S_n$ & number of returned samples (for source $n$) & Sec.~\ref{sec:methods-rejection} \\
    $\mathcal{G}$, $\mathcal{G}_{\rm max}$ & marginal likelihood of a prior sample, and its
      largest value over the prior samples
      & Secs.~\ref{sec:methods-marginal}, \ref{sec:methods-rejection} \\
    $J_{\rm eff}$ & effective sample size & Eq.~\ref{eq:ess} \\
    $\mathcal{H}$ & information (Kullback--Leibler divergence from prior to posterior)
      & Eq.~\ref{eq:information} \\
    \hline
    \multicolumn{3}{l}{\textbf{Population inference}} \\
    $\vec{\alpha}$, $\vec{\alpha}_0$ & hyperparameters of the population model and of the
      interim prior & Sec.~\ref{sec:intro} \\
    $Z_{{\rm int},n}$ & evidence for source $n$ under the interim prior
      & Eq.~\ref{eq:reweight} \\
    $w_{nk}$, $r_{nk}$ & weight and prior ratio of sample $k$ of source $n$
      & Eq.~\ref{eq:reweight} \\
    \hline
  \end{tabular}
\end{table*}

Below, we summarize the conventions that \harv\ uses for orbital phase, the orientation
of the orbit, and the signs of the amplitudes in the radial velocity and astrometric
models.
We choose these conventions so that a single set of orbital elements $(P, e, \tau,
\omega, \Omega, i)$ describes both the radial velocity and the astrometric orbit of an
observed star, so we can jointly model the two data types (e.g.,
Section~\ref{sec:cases-gaia4}).

Internally, we assume that times have been converted to a solar-system barycentric
reference frame, and we enforce that all datasets for a given source share a single
reference time $t_{\rm ref}$ (by default, the mean of all observation times).
The phase parameter $\tau \in [0, 1)$ is
the time of pericenter passage relative to the reference time in units of the period,
$t_{\rm peri} = t_{\rm ref} + \tau\,P$, so the mean anomaly is
\begin{equation}
  M(t) = \frac{2\pi}{P}\,(t - t_{\rm ref}) - 2\pi\,\tau
  \quad ,
\end{equation}
and the eccentric anomaly $E$ and true anomaly $f$ follow from Kepler's equation, $M = E
- e\,\sin E$.
The mean longitude used by the \texttt{SqrtEccRV} parameterization
(Table~\ref{tab:parameterizations}) is the value of $\omega + M$ at the reference time,
$\lambda = \omega - 2\pi\,\tau$.

In the orbital plane, we use Cartesian coordinates $(x', y', z')$ with $x'$ pointing
toward pericenter, $y'$ orthogonal to $x'$ and ahead in the direction of orbital
motion, and $z'$ perpendicular to the orbital plane, so that the position of the star
relative to the barycenter is
\begin{equation}
  \vec{r}' =
  \begin{pmatrix} x' \\ y' \\ z' \end{pmatrix} =
  r \begin{pmatrix} \cos f \\ \sin f \\ 0 \end{pmatrix} =
  a \begin{pmatrix} X \\ Y \\ 0 \end{pmatrix}
  \quad ,
  \label{eq:conv-orbit}
\end{equation}
where $r = a\,(1 - e^2) / (1 + e\,\cos f)$, and $X = x'/a = \cos E - e$ and $Y = y'/a =
\sqrt{1 - e^2}\,\sin E$ are the dimensionless orbital-plane coordinates that appear in
Equation~\ref{eq:al-model} and $a$ is the semi-major axis of the star's orbit about the
barycenter in length units.

On the sky, i.e., relative to the tangent plane at a given sky position, we use the same
coordinate frame as used by \gaia\ \citep{Halbwachs:2023}: $x$ points North (increasing
declination), $y$ points East (increasing right ascension), and $z$ points along the
line of sight \emph{away} from the observer, so that $z$ is the distance of the star
behind the system's barycenter (note that this frame is left-handed).
The two frames are related by a rotation via the standard orbital elements:\footnote{We
note that the orientation angles are defined such that the argument of pericenter
$\omega \in [0, 2\pi)$ and is the angle in the orbital plane from the ascending node to
the pericenter of the \textit{observed star's} orbit about the barycenter, the longitude
of the ascending node $\Omega \in [0, 2\pi)$ is the position angle of the ascending node
measured from North (i.e., it is the node at which the star crosses the plane $z=0$
moving away from the observer $\dot{z} > 0$), and the inclination $i \in [0, \pi]$ is
the angle between the orbital angular momentum vector and the line of sight (i.e., $i=0$
is face-on with angular momentum pointing towards the observer and the orbit moves
counter-clockwise on the sky, $i=\pi/2$ is edge-on, and $i=\pi$ is ``back''-on with
angular momentum pointing away from the observer and the orbit moves clockwise on the
sky).} longitude of ascending node, $\Omega$, inclination, $i$, and argument of
pericenter, $\omega$,
\begin{equation}
  \vec{r} =
  \begin{pmatrix} x \\ y \\ z \end{pmatrix} =
  \mat{R}\,\vec{r}'
  \quad ,
\end{equation}
where
\begin{align}
  \mat{R} &= \mat{R}_z(\Omega)\,\mat{R}_x(i)\,\mat{R}_z(\omega) \\
  \mat{R}_z(\phi) &=
    \begin{pmatrix}
      \cos\phi & -\sin\phi & 0 \\
      \sin\phi & \cos\phi & 0 \\
      0 & 0 & 1
    \end{pmatrix} \quad , \quad
  \mat{R}_x(\phi) =
    \begin{pmatrix}
      1 & 0 & 0 \\
      0 & \cos\phi & -\sin\phi \\
      0 & \sin\phi & \cos\phi
    \end{pmatrix}
  \quad .
\end{align}
Since $z' = 0$, only the first two columns of $\mat{R}$ contribute, and the position of
the star in the sky frame can also be written using the Thiele--Innes constants
\citep[e.g.,][]{Wright:2009, Halbwachs:2023}
\begin{equation}
  \begin{pmatrix} x \\ y \\ z \end{pmatrix} =
  \begin{pmatrix} A & F \\ B & G \\ C & H \end{pmatrix}
  \begin{pmatrix} X \\ Y \end{pmatrix}
  \quad ,
  \label{eq:conv-sky}
\end{equation}
where the six constants are $a$ times the elements of the first two columns of the
rotation matrix $\mat{R}$.
The astrometric Thiele--Innes constants,
\begin{align}
  A &= a\,(\cos\omega\,\cos\Omega - \sin\omega\,\sin\Omega\,\cos i) \nonumber \\
  B &= a\,(\cos\omega\,\sin\Omega + \sin\omega\,\cos\Omega\,\cos i) \nonumber \\
  F &= -a\,(\sin\omega\,\cos\Omega + \cos\omega\,\sin\Omega\,\cos i) \nonumber \\
  G &= -a\,(\sin\omega\,\sin\Omega - \cos\omega\,\cos\Omega\,\cos i)
  \quad ,
  \label{eq:conv-thiele-innes}
\end{align}
are defined to be consistent with the definitions used for \gaia\ DR3
\citep{Halbwachs:2023} and have the same units as $a$ (e.g., au or mas).
Recall that $X$ and $Y$ in these expressions are dimensionless positions in the orbital
plane (Equation~\ref{eq:conv-orbit}).
The other two constants, $C$ and $H$,
\begin{equation}
  C = a\,\sin\omega\,\sin i
  \quad , \quad
  H = a\,\cos\omega\,\sin i
  \quad ,
\end{equation}
set the distance of the star behind the system's barycenter,
\begin{equation}
  z = C\,X + H\,Y = r\,\sin i\,\sin(\omega + f)
  \quad .
  \label{eq:conv-z}
\end{equation}
In these expressions, $a$ is the semi-major axis of the star's orbit about the
barycenter and could be expressed in length units (e.g., au) or angular units (e.g.,
mas) depending on the context (most often angular units for astrometry applications).

We instead write the astrometric model (Equation~\ref{eq:al-model}) in terms of
angular offsets on the sky,
\begin{equation}
  \Delta\delta = \frac{\varpi\,x}{1~\unit{au}}
  \quad , \quad
  \Delta\alpha^* = \Delta\alpha\,\cos\delta = \frac{\varpi\,y}{1~\unit{au}}
  \quad ,
\end{equation}
relative to some reference position $(\alpha_0, \delta_0)$, where $(1~\unit{au}) /
\varpi$ is the distance to the system.
We can therefore write the astrometric model in terms of the Thiele--Innes constants, as
is common, such that
\begin{equation}
  \begin{pmatrix} \Delta\delta \\ \Delta\alpha^* \end{pmatrix} =
  \frac{\varpi}{1~\unit{au}}
  \begin{pmatrix} A & F \\ B & G \end{pmatrix}
  \begin{pmatrix} X \\ Y \end{pmatrix}
  \quad .
  \label{eq:conv-sky-angular}
\end{equation}
If we model the photocenter motion of a luminous star with a dark companion, then we
use the semi-major axis of the photocenter orbit, $a_0$, in place of $a$ in
Equation~\ref{eq:conv-thiele-innes}.
We note that $\Delta\alpha^*$, the right ascension offset $\alpha_0$, and the proper
motion $\mu_\alpha$ in Equation~\ref{eq:al-model} all include the factor $\cos\delta$
(these are often written $\alpha^*_0$ and $\mu_{\alpha^*}$).
The scan angle $\psi$ in Equation~\ref{eq:al-model} is the position angle of the
along-scan direction, measured from North through East (as provided by \gaia).

The radial velocity of the star relative to the systemic velocity $v_0$ is the $z$
component of the velocity, $\dot{z}$, so positive values mean the star is moving away
from the observer (i.e., redshift).
If we take the derivative of Equation~\ref{eq:conv-z} with respect to time, we get the
radial velocity model in Equation~\ref{eq:rv-model},
\begin{equation}
  {\rm RV}(t) = K\,[\cos(\omega + f) + e\,\cos\omega] + v_0
\end{equation}
with
\begin{equation}
  K = \frac{2\pi\,a\,\sin i}{P\,\sqrt{1 - e^2}}
  \quad ,
  \label{eq:conv-K}
\end{equation}
where again $a$ is the semi-major axis of the observed star's orbit about the
barycenter.
Again, if the companion is not luminous, the photocenter traces the observed star's
orbit, so $a_0 / \varpi = a / (1~\unit{au})$, and Equation~\ref{eq:conv-K} becomes
\begin{equation}
  K = \frac{2\pi\,\sin i}{P\,\sqrt{1 - e^2}} \, \frac{a_0}{\varpi} \, (1~\unit{au})
  \quad .
  \label{eq:conv-tie}
\end{equation}

Finally, we note that for typical runs, the default prior pdfs on $K$ and $a_0$ are
zero-mean (untruncated) normal distributions, so both amplitudes can take either sign.
The models have exact symmetries with this sign flip, because the radial velocity model
is invariant when $(\omega, K) \to (\omega + \pi, -K)$, and the astrometric model is
invariant when $(\omega, a_0) \to (\omega + \pi, -a_0)$ and $(\Omega, a_0) \to (\Omega +
\pi, -a_0)$.
However, for the astrometric model, applying both, $(\omega, \Omega) \to (\omega + \pi,
\Omega + \pi)$ means that the Thiele--Innes constants are not changed, but it does
change the sign of $z$ (Equation~\ref{eq:conv-z}) and therefore the radial velocity.
So, astrometry alone does not provide enough information to determine which of the two
nodes is the ascending node, which is why astrometric orbit catalogs typically restrict
$\Omega$ to $[0, \pi)$ \citep[e.g.,][]{Halbwachs:2023}.
We instead keep $\Omega \in [0, 2\pi)$, because radial velocity data break this
symmetry.
In \harv, the returned posterior samples have a method, \texttt{Samples.wrap\_angles()},
that maps posterior samples to the canonical representation with $K \geq 0$, $a_0 \geq
0$, and all angles in $[0, 2\pi)$, by first shifting $\omega$ by $\pi$ for samples with
$K < 0$ (or with $a_0 < 0$, if there are no radial velocity data), and then shifting
$\Omega$ by $\pi$ for samples where $a_0$ is still negative.

For double-lined (SB2) binaries, by convention, both components (primary 1 and secondary
2) of the system  share $(P, e, \tau, \omega)$, where $\omega$ is the argument of
pericenter of the primary.
Since the secondary is on the opposite side of the barycenter, its orbit is technically
at $\omega + \pi$, but we instead keep the same $\omega$ for both components and force
$K_2$ to be negative.

\section{Marginalizing over linear parameters with truncated prior pdfs}
\label{apx:truncated}

Here we derive the expressions referenced in Section~\ref{sec:methods-marginal} for the
marginal likelihood when the prior pdf over one or two of the linear parameters is a
truncated normal.
We also describe how we evaluate these expressions in the code, and how we then generate
posterior samples of the linear parameters when they have truncated priors.

We write a truncated normal prior pdf over parameters $\vec{\beta}$, which may depend on
other parameters $\vec{\theta}$, as
\begin{equation}
  p(\vec{\beta} \given \vec{\theta}) = \frac{1}{Z_{\rm prior}} \,
    \norm(\vec{\beta} \given \vec{\mu}, \mat{\Lambda}) \, \mathbf{1}_\mathbb{S}(\vec{\beta})
  \quad ,
\end{equation}
where the indicator function $\mathbf{1}_\mathbb{S}(\vec{\beta})$ is equal to one when
$\vec{\beta}$ is in the allowed domain $\mathbb{S}$ and zero otherwise, and
$Z_{\rm prior}$ is the normalization constant that makes the pdf integrate to 1 over
$\mathbb{S}$ (so here, $\mathcal{N}(\cdot)$ is the untruncated normal pdf).
The allowed domain is rectangular, so each truncated parameter
$\beta_{k'}$ is restricted to an interval $(l_k, h_k)$, and is potentially one-sided, in
which either $l_k$ or $h_k$ may be infinite.
The other linear parameters $\beta_{k\neq k'}$ are unconstrained.
The marginal likelihood is the integral of the likelihood times this prior pdf.
Using the refactorization in Equation~(\ref{eq:refactor}),
\begin{eqnarray}
  p(\vec{y} \given \vec{\theta})
  &=& \frac{1}{Z_{\rm prior}} \int \dd\vec{\beta} \,
    \norm(\vec{y} \given \mat{X}\,\vec{\beta}, \mat{C}) \,
    \norm(\vec{\beta} \given \vec{\mu}, \mat{\Lambda}) \, \mathbf{1}_\mathbb{S}(\vec{\beta}) \\
  &=& \frac{1}{Z_{\rm prior}} \,
    \norm(\vec{y} \given \mat{X}\,\vec{\mu},
      \mat{C} + \mat{X}\,\mat{\Lambda}\,\transp{\mat{X}})
    \int_{\mathbb{S}} \dd\vec{\beta} \, \norm(\vec{\beta} \given \vec{a}, \mat{A}) \\
  &=& \frac{Z_{\rm post}}{Z_{\rm prior}} \,
    p_{\rm untrunc}(\vec{y} \given \vec{\theta})
  \quad ,
\end{eqnarray}
which is Equation~(\ref{eq:truncated}).
The second expression follows from the fact that the first Gaussian on the right-hand
side of Equation~(\ref{eq:refactor}) does not depend on $\vec{\beta}$.

We assume that our prior pdfs over the linear parameters are independent, so
$\mat{\Lambda}$, the linear parameter prior covariance matrix, is diagonal with elements
$\sigma_k^2$, and $Z_{\rm prior}$ is thereforea product over the truncated parameters,
\begin{equation}
  Z_{\rm prior} = \prod_k \left[
    \Phi\!\left(\frac{h_k - \mu_k}{\sigma_k}\right)
    - \Phi\!\left(\frac{l_k - \mu_k}{\sigma_k}\right) \right]
  \quad ,
\end{equation}
where $\Phi$ is the standard normal cumulative distribution function (cdf).
This is valid for any number of truncated parameters.

Unfortunately, in general, the covariance matrix $\mat{A}$ of the conditional posterior
pdf is not diagonal, so $Z_{\rm post}$ does not factorize like the prior.
However, it does simplify because we integrate the unconstrained parameters over their
full domain, and marginalizing a Gaussian over some of its dimensions leaves a Gaussian
in the remaining dimensions with the corresponding elements of the mean and covariance
matrix.
$Z_{\rm post}$ is therefore the integral over the truncated domain of a Gaussian in
only the truncated parameters.
For each truncated parameter, we define the standardized bounds
\begin{equation}
  \tilde{l}_k = \frac{l_k - a_k}{\sqrt{A_{kk}}}
  \quad , \quad
  \tilde{h}_k = \frac{h_k - a_k}{\sqrt{A_{kk}}}
  \quad .
\end{equation}
With only one truncated parameter,
$Z_{\rm post} = \Phi(\tilde{h}_k) - \Phi(\tilde{l}_k)$, i.e., a difference of the cdf of
the standard normal evaluated at the bounds.
With two truncated parameters, which we label 1 and 2, the Gaussian posterior over only
the truncated parameters is specified by the standardized bounds and the correlation
coefficient $\rho = A_{12} / \sqrt{A_{11}\,A_{22}}$.
The integral over the rectangular truncated domain is then a sum over the four corners,
\begin{equation}
  Z_{\rm post} =
    \Phi_2(\tilde{h}_1, \tilde{h}_2; \rho)
    - \Phi_2(\tilde{l}_1, \tilde{h}_2; \rho)
    - \Phi_2(\tilde{h}_1, \tilde{l}_2; \rho)
    + \Phi_2(\tilde{l}_1, \tilde{l}_2; \rho)
  \quad ,
  \label{eq:corners}
\end{equation}
where $\Phi_2(u, v; \rho)$ is the cdf of the standard bivariate normal distribution with
correlation coefficient $\rho$.
When a bound is infinite the corresponding terms simplify, because
$\Phi_2(u, v; \rho) = 0$ if either argument is $-\infty$, and
$\Phi_2(u, +\infty; \rho) = \Phi(u)$.

There is no closed-form expression for $\Phi_2$, but we can write it as
$\Phi(u)\,\Phi(v)$ plus a one-dimensional integral over the correlation coefficient
(from 0 to $\rho$) \citep{Plackett:1954}.
Through a change of variables, substituting $\rho \rightarrow \sin\xi$, we can transform
the integral into a form that is bounded and smooth, so we evaluate the integral with a
fixed Gauss--Legendre quadrature with 64 nodes over the interval $[0, \arcsin\rho]$.
With fixed bounds and number of nodes, we can compile this evaluation with \jax\ and
autodiff works through this as expected.

There is one subtlety with implementing this approach numerically, which is that the sum
in Equation~(\ref{eq:corners}) can lose numerical precision when the truncated domain
contains a small fraction of the probability mass of the conditional posterior.
In this case, the four terms in Equation~(\ref{eq:corners}) nearly cancel, so the result
is dominated by round-off error in the sum.
This issue is worse when $Z_{\rm post}$ is small, which occurs when the truncated domain
is far from the mean of the conditional posterior.
We clamp the value of $Z_{\rm post}$ so that it is set to $Z_{\rm post}$ = 0 when the
value is too small to be distinguished from round-off error in the sum.
This technically under-weights some affected samples, but the alternative (not clamping)
can lead to a value that over-weights the samples, which is worse.
In practice, this only discards samples whose weights would have been negligible anyway.

When three or more linear parameters have truncated domains, $Z_{\rm post}$
is an integral of a three or more dimensional Gaussian with a dense covariance matrix
over a rectangular domain.
There is no simple expression for this integral, so we do not support this case.

Given a sample of the nonlinear parameters $\vec{\theta}$, we need to generate a sample
of the linear parameters from the truncated conditional posterior pdf,
$\norm(\vec{\beta} \given \vec{a}, \mat{A})$.
With one truncated parameter, we sample its value from a one-dimensional truncated
normal with mean $a_k$ and variance $A_{kk}$, using \package{numpyro}.
There is no equivalent in \package{numpyro} for two correlated truncated parameters, so
we draw $\beta_1$ with inverse transform sampling.
The cdf of the conditional posterior pdf over $\beta_1$ within the truncated domain is
\begin{equation}
  F(t) = \frac{1}{Z_{\rm post}} \left[
    \Phi_2(\tilde{t}, \tilde{h}_2; \rho)
    - \Phi_2(\tilde{l}_1, \tilde{h}_2; \rho)
    - \Phi_2(\tilde{t}, \tilde{l}_2; \rho)
    + \Phi_2(\tilde{l}_1, \tilde{l}_2; \rho) \right]
  \quad ,
\end{equation}
with $\tilde{t} = (t - a_1)/\sqrt{A_{11}}$.
This is equivalent to Equation~(\ref{eq:corners}) with the upper bound $h_1$ replaced by
$t$, divided by $Z_{\rm post}$.
We draw $u$ uniformly between 0 and 1 and solve $F(\beta_1) = u$ with bisection using a
fixed number of steps (60), which, in our tests, resolves $\beta_1$ sufficiently
precisely at double precision.
We then draw $\beta_2$ given $\beta_1$ from a one-dimensional truncated normal with mean
$a_2 + A_{12}\,(\beta_1 - a_1)/A_{11}$ and variance $A_{22} - A_{12}^2/A_{11}$.
We sample the unconstrained parameters from their Gaussian pdf conditioned on both of
the truncated parameters.

\bibliographystyle{AASTeX7/aasjournalv7}
\bibliography{refs}{}

\end{document}